\documentclass[twocolumn,aps,showpacs,showkeys,prb,superscriptaddress,reprint]{revtex4-1}
\usepackage{graphicx}
\usepackage{amsmath}
\usepackage{amssymb}
\usepackage{amsfonts}
\usepackage{color}
\usepackage{bm}% bold math
\usepackage{hyperref}

\hypersetup{
	colorlinks = true,
	pdftitle = {},
	pdfauthor = {},
	pdfkeywords = {},
	linkcolor = blue,
	citecolor = blue,
	filecolor = black,
	urlcolor = magenta
}

\newcommand{\lz}{\langle \hat{L}_z \rangle}
\newcommand{\qmax}{q_\text{max}}
\newcommand{\Vacc}{V_\text{acc}}
\newcommand{\datt}{d_\text{at.col}}

\begin{document}

%\title{Computational optimization of magnetic measurements with electron vortex beams at atomic resolution}
\title{Scattering of electron vortex beams on a magnetic crystal: towards atomic resolution magnetic measurements}

\author{J\'{a}n Rusz}
\affiliation{Department of Physics and Astronomy, Uppsala University, P.O. Box 516, 75120 Uppsala, Sweden}
%\affiliation{Institute of Physics, Czech Academy of Sciences, Na Slovance 2, CZ-182 21 Prague, Czech Republic}
\author{Somnath Bhowmick}
\affiliation{Department of Materials Science and Engineering, Indian Institute of Technology, Kanpur 208016, India}
\author{Mattias Eriksson}
\author{Nikolaj Karlsson}
%\author{Olle Eriksson}
\affiliation{Department of Physics and Astronomy, Uppsala University, P.O. Box 516, 75120 Uppsala, Sweden}

\begin{abstract}
Use of electron vortex beams (EVB), that is convergent electron beams carrying an orbital angular momentum (OAM), is a novel development in the field of transmission electron microscopy. They should allow measurement of element-specific magnetic properties of thin crystals using electron magnetic circular dichroism (EMCD)---a phenomenon similar to the x-ray magnetic circular dichroism. Recently it has been shown computationally that EVBs can detect magnetic signal in a scanning mode only at atomic resolution. In this follow-up work we explore in detail the elastic and inelastic scattering properties of EVBs on crystals, as a function of beam diameter, initial OAM, acceleration voltage and beam displacement from an atomic column. We suggest that for a 10~nm layer of bcc iron oriented along (001) zone axis an optimal configuration for a detection of EMCD is an EVB with OAM of $1\hbar$ and a diameter of 1.6~\AA, acceleration voltage 200~keV and an annular detector with inner and outer diameters of $G$ and $5G$, respectively, where $\mathbf{G}=(100)$.
\end{abstract}


\maketitle

\section{Introduction}

Nanostructures involving magnetic materials are used in diverse applications, such as data storage, chemical catalysis, medical diagnostics and treatment, or removal of heavy elements from waste water. Fine-tuning their properties demands methods capable of quantitative magnetic characterization at lateral resolutions from nano-scale down to a few atoms. One of such candidate measurement methods is electron magnetic circular dichroism (EMCD; \cite{nature}), a spectroscopic experiment performed using a transmission electron microscope (TEM).

EMCD as an experimental technique is known for about 10 years. Since its inception \cite{emcdproposal} it went through a rapid development with significant improvements in spatial resolution and signal to noise ratio \cite{lacbed,lacdif,emcd2nm,lsfollow,polyemcd}. Theoretical studies provided understanding of the interplay of dynamical diffraction effects and inelastic excitations that give rise to EMCD signal \cite{prbtheory,bwconv} and its relation to the ground state expectation values \cite{opmaps} such as spin and orbital magnetic moments via sum rules \cite{oursr,lionelsr}. Although early adopters have already successfully applied this characterization technique in their research \cite{nanostuff,klie,bacteria,nanozno,nanofe3o4,cro2,fe3o4chan}, yet, EMCD has not reached a stage of wide adoption as a routine experimental method. The reasons are two-fold: 1) requirement of single-crystalline specimen precisely oriented in a two-beam or three-beam orientation, and 2) generally EMCD spectra have often a low signal to noise ratio, which is due to the fact that EMCD needs to be measured aside the Bragg spots.

Several approaches have been suggested to overcome the requirement of single-crystalline specimen and specific orientations. An approach using a Boersch plate \cite{boersch47} was suggested by Hasenkopf et al.\cite{boersch,diploma_hasenkopf}. A convergent beam should be split into two parts, one of which should be phase-shifted by $\frac{\pi}{2}$ using a coil in the Boersch plate. These two beams would be focused back on the same spot on the sample and via a coherent interference and dynamical diffraction, an EMCD signal should appear at Thales circle positions\cite{emcdproposal}. To this date, this approach has not been yet successfully implemented.

Promising new approach is based on an application of advanced statistical methods to extract the EMCD signal, previously theoretically explored in Ref.~\cite{mcremcd}. In this experiment one acquires a large number of core-level spectra in more-or-less random geometrical conditions (mutually random beam, sample and detector orientations) and the magnetic and nonmagnetic signal components are separated by a statistical post-processing. This method was recently successfully applied to a polycrystalline iron film \cite{polyemcd}.

Finally, a method based on utilizing electron vortex beams\cite{vorttem} (EVBs) has been suggested\cite{vortjo}. EVBs are typically convergent electron beams that carry a nonzero orbital angular momentum (OAM). By measuring an electron energy loss spectrum once using EVB with $\lz=+1\hbar$ and another one with $\lz=-1\hbar$, a nonzero EMCD signal should be obtained as their difference. Because the measurement of the spectrum is done at the transmitted beam, it means significantly stronger intensity (assuming that we can obtain EVB with an intensity comparable to an ``ordinary'' convergent electron beam). Originally it was believed that this method will allow to measure EMCD without a requirement of single crystals in specific orientations. Follow-up theoretical works discussed the elastic propagation of EVBs via vacuum \cite{vortexfree} or crystals \cite{vortcryst,lubk,xin}, their inelastic interaction with atoms \cite{lloydprl,lloydpra,yuan} or formation \cite{idrobo}. Experimental works provided new methods of generating EVBs \cite{mcmorran,saitoh2,strongvortex,clarkaberr,needleUM,monopole}, reducing their diameter \cite{vortatom,vortat2}, or measuring their angular momentum.\cite{saitoh} Yes, new measurement of EMCD have not appeared in the literature so far.

In our recent work we have computationally shown that, in fact, utility of EVBs for measurement of EMCD is limited to measurements at an atomic resolution \cite{vortexelnes}. If a thin crystalline layer is oriented in a low index zone axis, then a narrow EVB passing through an atomic column will be sensitive to its magnetic properties. However, if the beam passes in between columns or if its diameter is significantly larger than interatomic spacing, the EMCD signal won't be detected at the transmitted beam anymore. Very recently, detection of EMCD was ruled out for nanoparticles larger than 1~nm\cite{schattnp}.

In this work we describe our implementation of a theory of inelastic electron scattering for EVB, which is based on a combination of multislice propagation method for the incoming beam and Bloch-waves (BW) description for the outgoing beam. The method is built on top of the efficient summation algorithm \textsc{mats}\cite{bwconv} introduced recently. \textsc{mats} algorithm performs an efficient filtering and summation of the largest terms in the Bloch-waves expression of the double-differential scattering cross-section, containing an 8-fold sum over Bloch coefficients.\cite{kohl,rossouw,saldin} Using this method we explore in detail the inelastic scattering of EVBs on crystals as a function of EVB diameter, acceleration voltage, starting angular momentum and displacement from an atomic column. Finally, we propose optimized settings for an EMCD experiment with vortex beams.

%In the Section~\ref{sec:elast} we discuss the multislice simulation of the elastic propagation of the EVB through the sample with focus on exchange of angular momentum between the EVB and sample. In subsection \ref{sec:symmetry} we discuss the symmetry properties of the system consisting of the beam with an angular momentum and the crystal with its own rotation and mirror symmetries. In section~\ref{sec:inel} we first describe our computational approach that combines the multislice and Bloch-waves methods. The results of the simulations are then summarized in following subsections.

\section{Survey of the parameter space}

\begin{table*}[tb]
 \begin{tabular}{lcl}
   \hline \hline
   Parameter & Notation & Values \\
   \hline
   Acceleration voltage & $\Vacc$ (kV)  & 50, 100, 200, 300, 500, 1000 \\
   Starting OAM         & $\lz$ ($\hbar$)      & 0, 1, 2, 3, 5, 8, 10 \\
   Probe size           & $\qmax$ (a.u.$^{-1}$) & 0.1, 0.2, 0.3, 0.4, 0.5, 0.6, 0.7, 0.8, 0.9, 1.0 \\
   Distance from at. column & $\datt$ (\AA{}) & 0.0, 0.137, 0.342, 0.683, 1.435 \\
   Sample thickness     & $t$ (nm)          & 10.04, 20.09, 30.13, 40.18 \\
   \hline \hline
 \end{tabular}
 \caption{Explored parameter space.}
 \label{tab:params}
\end{table*}

We have performed a systematic study of the elastic and inelastic scattering of EVBs on bcc iron crystal as function of a set of several independent parameters. The following parameters have been considered: acceleration voltage, starting angular momentum, probe size, distance from the atomic column and sample thickness. Considered values of these parameters are summarized in Table~\ref{tab:params}.
%The vast available dataset can hardly be explored in its entirety and will likely stimulate further research with foci on specific situations.

\begin{figure}
  \includegraphics[width=8.5cm]{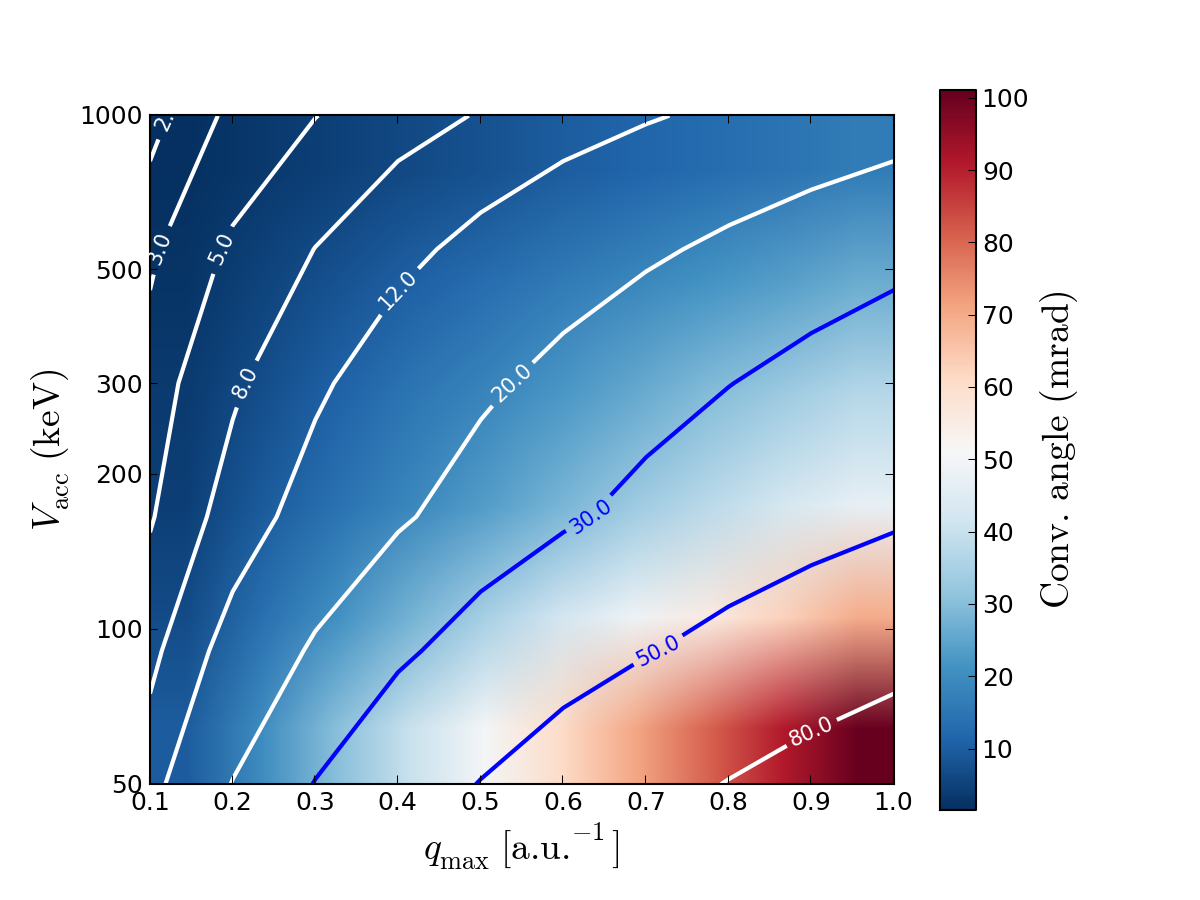}
  \caption{Contour plot showing the convergence angle (mrad) needed to achieve disc of radius $\qmax$ (a.u.$^{-1}$) at given acceleration voltage (kV). Some of the values of the convergence angles, particularly for low voltages and large $\qmax$, are beyond the range of feasible convergence angles without extensive aberrations.}
  \label{fig:convqmax}
\end{figure}

\begin{figure}
  \includegraphics[width=8.5cm]{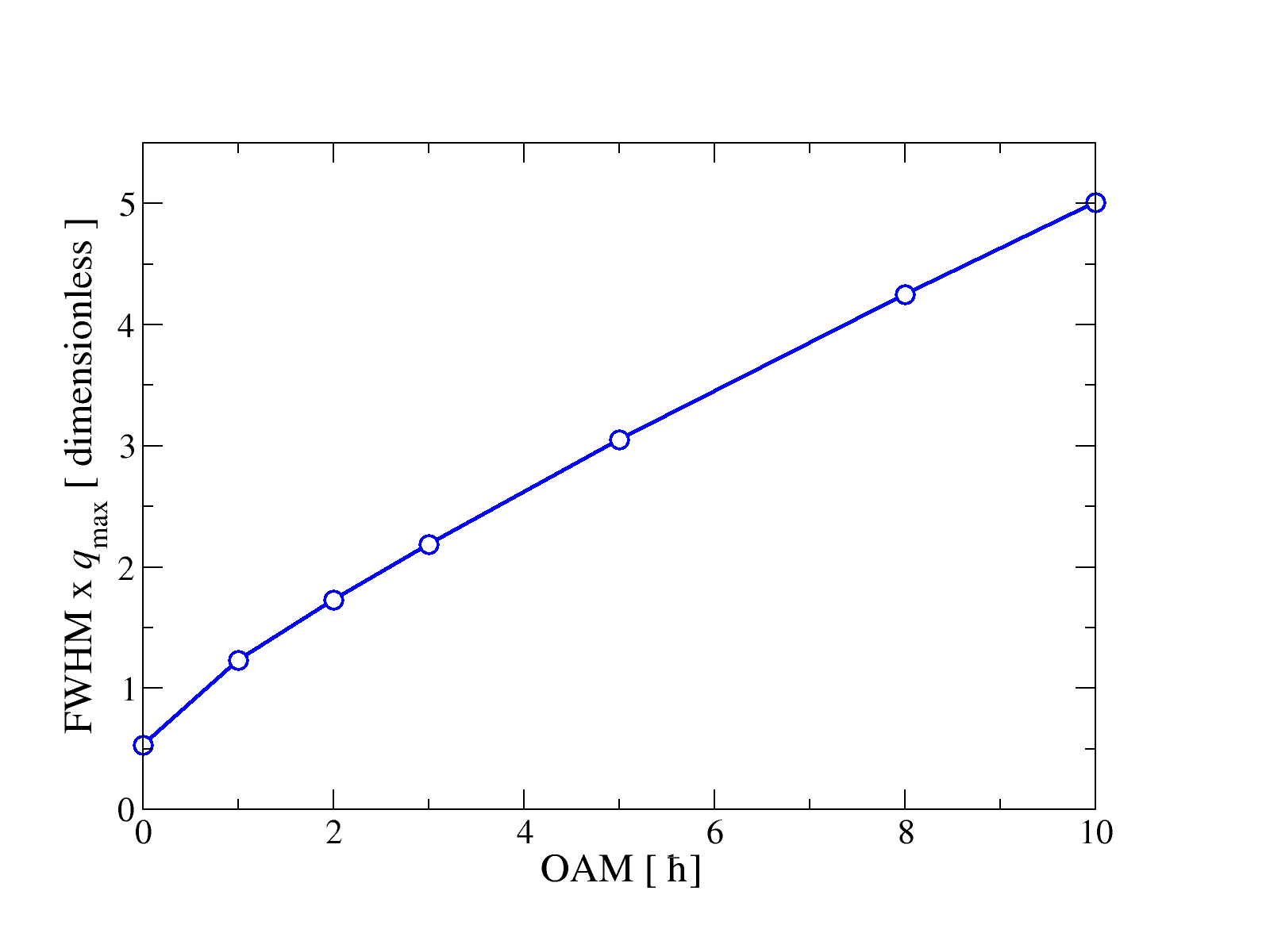}
  \caption{Dependence of the FWHM beam diameter on the angular momentum. Displayed is an universal curve FWHM multiplied by $\qmax$ (see text for details).}
  \label{fig:fwhmxqmax}
\end{figure}

The probe size parametrization deserves a more detailed explanation. It has been expressed in terms of the radius of a disc describing the beam in $\mathbf{k}$-space, denoted $\qmax$. The convergence angle necessary to reach given $\qmax$ depends on the acceleration voltage, see Fig.~\ref{fig:convqmax}. An advantage of this choice is that the beam diameter, to which $\qmax$ translates, does not depend on the acceleration voltage, because the probe wave-function in the real space is obtained directly by Fourier transform of the disc. However, the beam diameter does depend on the OAM of the beam. The analytical expression for the vortex beam shape is given by \cite{vortexfree}
\begin{equation}\label{eq:vortexwf}
\psi(r,\phi) \propto \mathcal{FT}\left[ e^{im\phi_k} \Theta(\qmax-k) \right]
\end{equation}
which is a Fourier transform of a disc in $\mathbf{k}$-space of diameter $\qmax$, with phase of the $\mathbf{k}$-space wavefunction dependent on the azimuthal angle $\phi_k$ and OAM $\lz{}=m\hbar$. Based on this expression, for a given $\lz{}=m\hbar$ the resulting wave function shape in the real space is the same for all values of $\qmax$, up to a scaling. By multiplying the diameter of the probe (using the full-width at half-maximum, FWHM) with $\qmax$ we obtain a dimensionless measure of the beam diameter, which can be conveniently expressed as an universal function of the OAM, Fig.~\ref{fig:fwhmxqmax}. From this figure one can directly read the beam diameters in Bohr radii (atomic unit, 1~a.u.=0.529178~\AA) for $\qmax=1.0$ a.u.$^{-1}$, i.e., the largest value considered here. The beam diameters for an arbitrary $\qmax$ value are obtained by dividing the FWHM$\times \qmax$ with the $\qmax$ value itself.

Once a beam described by Eq.~\ref{eq:vortexwf} passes through a crystal, it will diffract and the resulting diffraction pattern will consist of several (possibly overlapping) discs. The technique of using convergent electron beams is called convergent beam electron diffraction (CBED) and the discs are often referred to as CBED discs.

\section{Elastic scattering of electron vortices\label{sec:elast}}

\subsection{Multislice code and computational parameters\label{sec:mult}}

Inelastic scattering simulations described in the next section are rather time consuming, therefore optimization of computational parameters is of great importance. Major parameters for the elastic multislice part of the simulation\cite{kirkland} are the sizes of the grids, both within a unit cell and the number of unit cells included in a simulation. We have adopted a grid of $42 \times 42 \times 42$ grid points within a single unit cell, which turned out to be sufficient to accurately reproduce a Bloch-waves calculation for a plane-wave illumination (see below). 

The target maximum thickness in our simulations was 40nm, which means approximately 140 unit cells of bcc iron ($a=2.87$\AA). The number of unit cells in the $x,y$ directions depends sensitively on the beam characteristics. For a plane wave $1a \times 1a$ is enough. But for convergent beams (with or without angular momentum) that can only rarely be sufficient---either the beam is too wide to fit into one unit cell, or its convergence angle is so large that the beam quickly spreads, as it propagates through the crystal. Beam characteristics in our simulations are determined by the $\qmax$ and $\lz$ parameters defined above. We have not considered the broadening of the beam due to finite source size, i.e., we assume purely diffraction-limited beam diameter. Naturally, small values of $\qmax$ mean a broad beam, which spreads very slowly, while large values of $\qmax$ produce very well focused beams, which however spread quickly as they propagate through the lattice. For actual simulations we have used supercell size $20a \times 20a$, that is a square with approximately 5.7nm long edge. The grid size is thus $840 \times 840$ pixels.

All the calculated 3-dimensional probe wavefunctions were stored and serve as an input for the calculations of the double-differential scattering cross-section described in Section~\ref{sec:inel}.

\subsection{Symmetry considerations\label{sec:symmetry}}

\begin{figure}[t]
  \includegraphics[width=8.6cm]{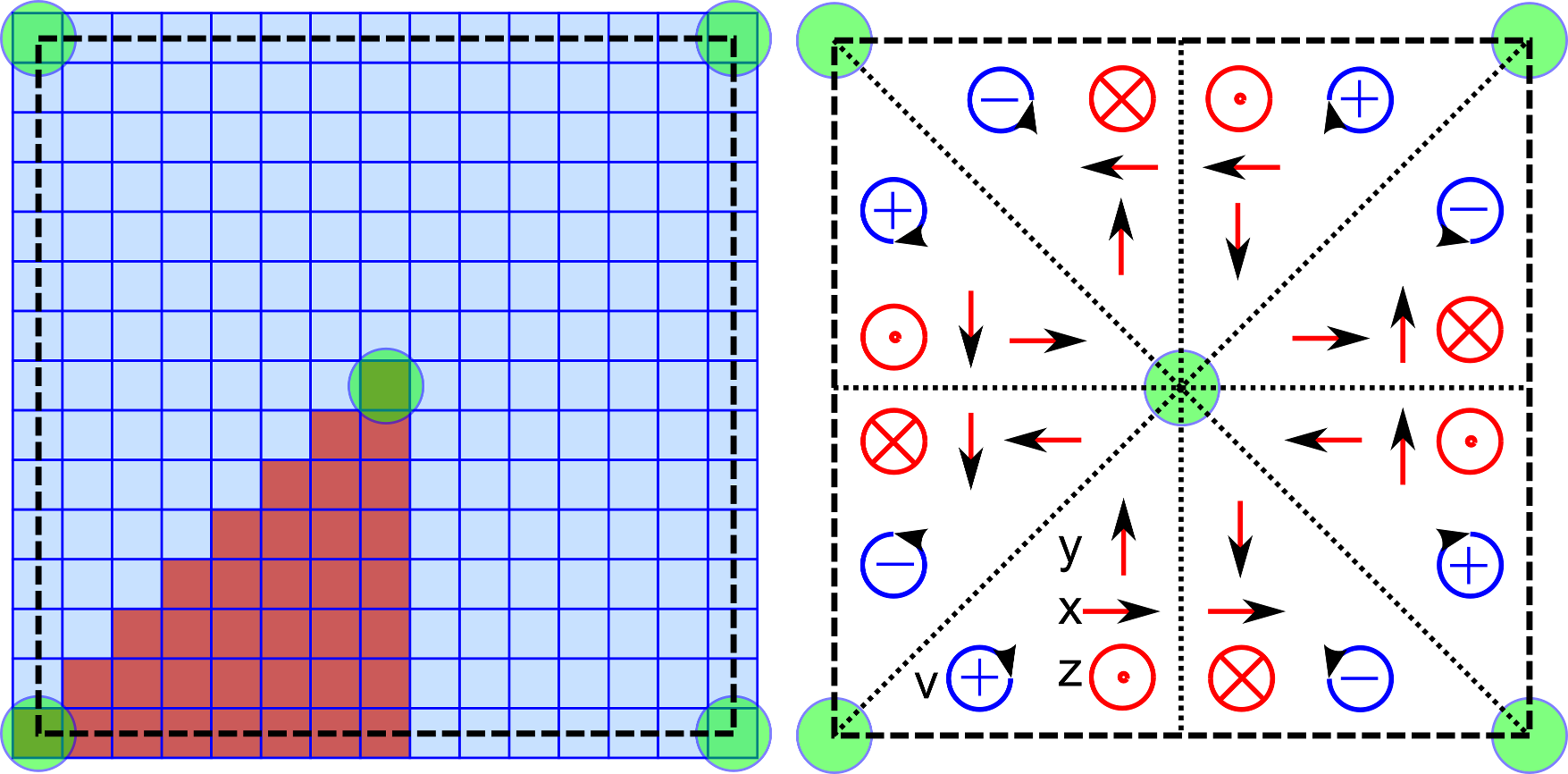}
  \caption{a) Schematic figure of beam centers within the unit cell considered in our simulations. Green circles mark the positions of atomic columns and the dashed line marks the border of a unit cell in $x,y$-directions. The red squares represent the beam centers. Symmetry operations allow to translate the 36 beam centers to their symmetric equivalents and fill the whole unit cell with a mesh of $14 \times 14$ pixels. b) Transformations of the magnetic moment direction (red; $x,y,z$) and the sign of the beam angular momentum (blue; $v$) after application of mirror or rotation symmetries. The picture demonstrates symmetry relations connecting the triangular wedges of the unit cell (see text for details).}
  \label{fig:positions}
\end{figure}

For calculations of the energy-filtered scans at atomic resolution it is advantageous to use symmetry properties of the crystal structure and relation between the EVBs with positive and negative angular momentum. If the magnetization of the sample points in $z$-direction, a rotation by 90 degrees along the $z$-axis is a symmetry operation for both the beam and the sample. That means that it is certainly enough to sample only one quarter of the unit cell (e.g., the lower left square of dimensions $\frac{a}{2}\times\frac{a}{2}$) and fill in the rest by rotating the square around the center of the cell. However, we can also utilize the diagonal mirror symmetry operation of the unit cell, if we take into account that a mirror will invert the sign of all axial vectors lying within its symmetry plane---that is both the magnetization vector and the angular momentum of the EVB. In other words, an EMCD signal calculated for an EVB with $\lz{}=+n\hbar$ with vortex core positioned at $(x,y)$ will have an opposite sign to an EMCD calculated for an EVB with $\lz{}=-n\hbar$ passing through $(y,x)$.
%the $(x,y)$ calculated for a vortex beam with angular momentum $+n\hbar$ will have an equal value to a point $(y,x)$ calculated for a beam with angular momentum $-n\hbar$ and reversed magnetization direction.

Situation is somewhat different, if magnetization is in-plane, let's say along $x$-direction. In such case, a rotation by 90 degrees along the $z$-axis changes the magnetization vector to plus or minus $y$-direction. The action of mirror symmetry planes parallel with $x$ (or $y$) crystal axis will invert (or keep) the sign of magnetization. The diagonal axis combines these two effects - the component parallel with mirror axis gets inverted sign and the component perpendicular to the mirror axis stays unchanged. As a result, from magnetization along $x$ direction we obtain magnetization along plus or minus $y$ direction. We remind that all these mirror planes also invert the sign of EVB angular momentum.

Using these symmetry properties, it is enough to sample 1/8th of the unit cell area with beams of angular momenta $\pm \hbar$ and $0$ for all three directions of magnetization $x,y,z$, and the rest of the unit cell area can be obtained using symmetry operations. This is summarized by Fig.~\ref{fig:positions}. The left panel shows the grid of points within the unit cell, highlighting the pixels within a triangular area covering 1/8th of unit cell, for which the calculations were actually performed. On the right side of the figure all mirror symmetry operations are shown and their influence on the direction of magnetic moments along $x,y$ and $z$-direction (red color) and angular momentum of the beam (blue color).

\subsection{Orbital angular momentum\label{sec:orb}}

\begin{figure}[t]
  \includegraphics[width=8.6cm]{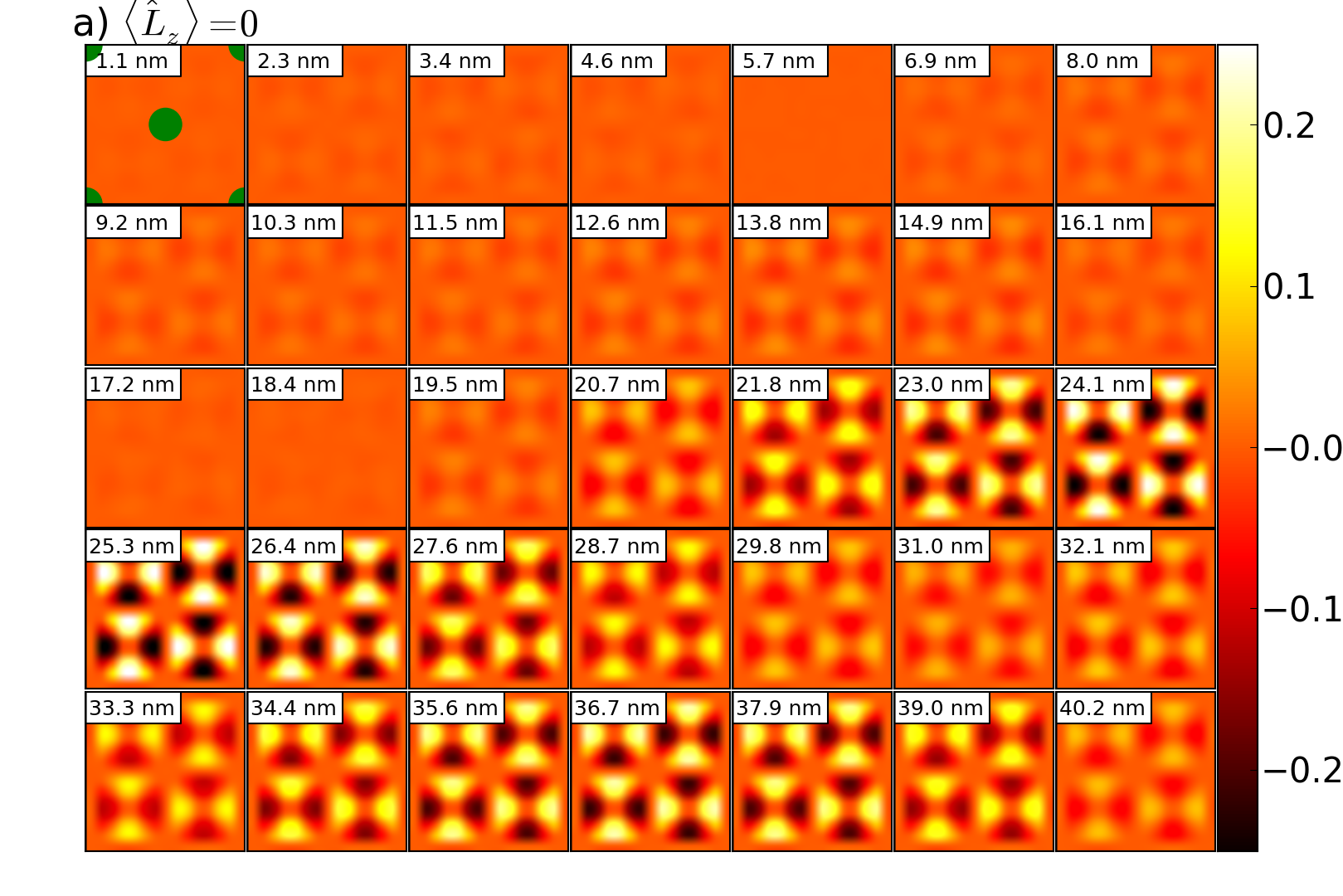}
  \includegraphics[width=8.6cm]{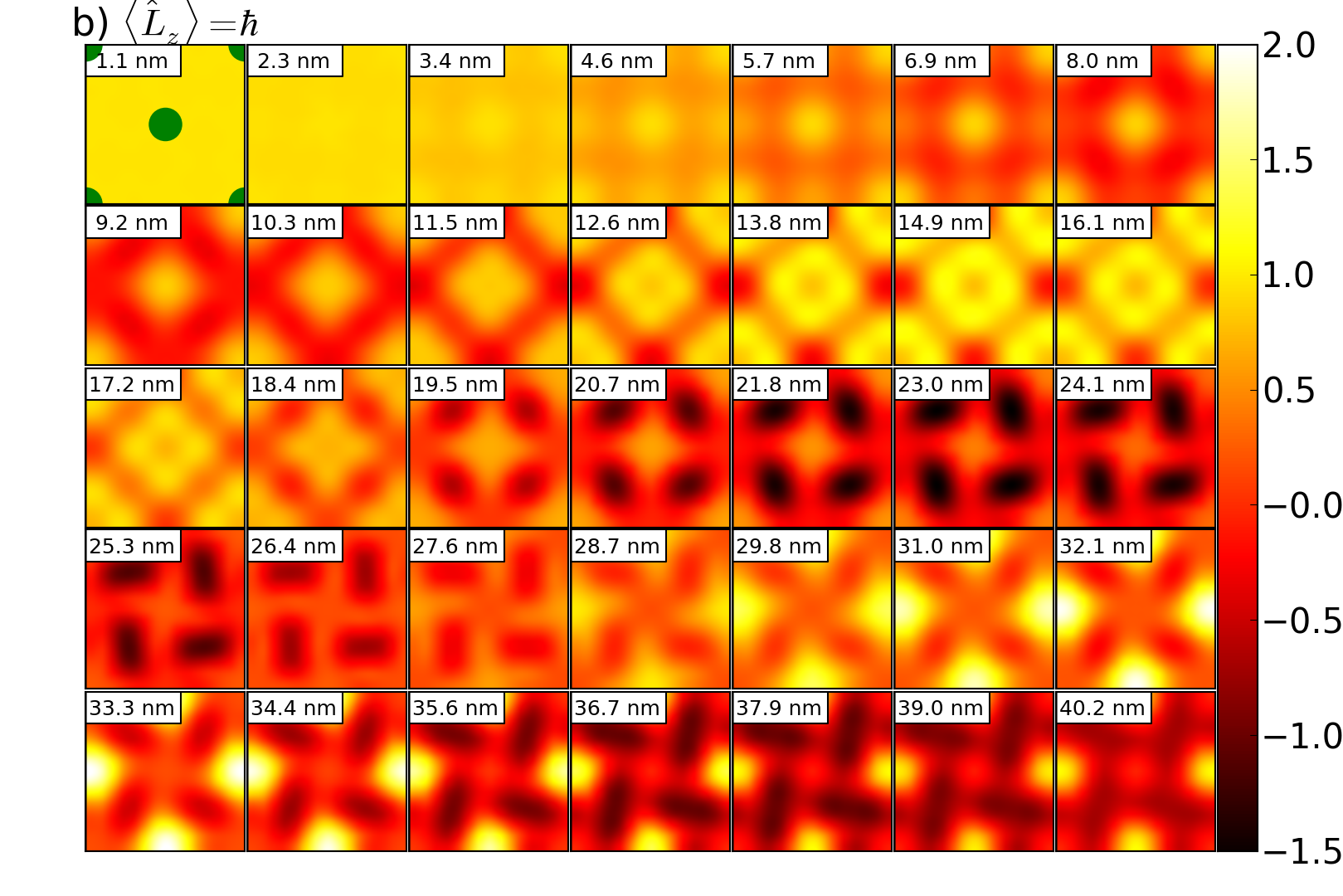}
  \caption{Evolution of the OAM of a narrow beam ($q_\text{max}=0.3$~a.u.$^{-1}$) as a function of sample thickness and position of the vortex core within a unit cell. Top and bottom panels correspond to a beam with angular momentum a) $\lz{}=0$ and b) $\lz{}=1\hbar$, respectively. Each square pattern maps shows a slice over a complete unit cell, thicknesses are in multiples of 4 unit cells, $4a \approx 1.15$nm. Green circles mark positions of the atomic columns.}
  \label{fig:orbmommap}
\end{figure}

We have performed a set of 36 calculations on a grid covering 1/8-th of an area of unit cell, Fig.~\ref{fig:positions}, one set for each of angular momenta $\lz{}=-\hbar,0,\hbar$. Here we considered $\qmax{}=0.3$ a.u.$^{-1}$ at 200 keV, a value which is in between the beam widths considered in our previous manuscript\cite{vortexelnes}. At these settings, FWHM of beam with zero angular momentum is 1.8~a.u.=0.95~\AA{} and beam with angular momentum $\pm \hbar$ has a diffraction limited FWHM of 4.1~a.u.=2.2~\AA, as can be extracted from Fig.~\ref{fig:fwhmxqmax}. The maps of the OAM as a function of sample thickness are summarized in Fig.~\ref{fig:orbmommap}.

The exchange of angular momentum between beam and lattice sensitively depends on the illumination spot. Even the beam with starting angular momentum of zero (Fig.~\ref{fig:orbmommap}, top panel) does acquire some angular momentum as it propagates through the lattice. At thicknesses beyond 10nm it acquires non-negligible angular momentum, peaking at around $0.25\hbar$ at thickness of 25nm. As can be seen, the patterns of angular momentum are highly symmetric and areas of positive angular momentum are matched in shape and amplitude by areas of negative angular momentum. As a result, an averaged value over the whole unit cell is zero at all thicknesses. On the other hand, these results indicate that for a sufficiently narrow beam, we can generate beam with nonzero angular momentum simply by passing it through a crystal of appropriate thickness at an appropriate position within the unit cell (assuming that such a task is or will be technically feasible).

A beam with starting OAM $\lz{}=1\hbar$ shows a different pattern, see Fig.~\ref{fig:orbmommap}, bottom panel. The symmetry has changed and, in general, the average over the unit cell does not vanish. As in the case of beam with $\lz{}=0$, by illuminating an appropriate spot in the lattice for a sample of suitable thickness it is possible to manipulate the probe's angular momentum. However, the range of accessible values is substantially enhanced compared to a probe with zero initial angular momentum.

\section{Inelastic scattering of electron vortex beams\label{sec:inel}}

\subsection{Theory\label{sec:theory}}

In this section we describe our approach to the evaluation of the dynamical diffraction effects and inelastic electron scattering for EVB. In fact, this method is applicable to illumination by an arbitrary coherent beam or a combination of beams, but here we restrict the treatment to the EVB with an arbitrary value of $\lz$. We will consider core-level excitations of $2p$ electrons of bcc iron into unoccupied $3d$ states, i.e., the $L_{2,3}$ edge transitions.

In the Bloch waves (BW) formulation, the double-differential scattering cross-section (DDSCS) is given by
\begin{eqnarray} \label{eq:dscsfin}
  \frac{\partial^2 \sigma}{\partial\Omega \partial E} & = & 
    \sum_{\substack{jlj'l'\\\mathbf{ghg'h'}}}  C_{\mathbf{0}}^{(j)\star} C_{\mathbf{g}}^{(j)}
                   D_{\mathbf{0}}^{(l)} D_{\mathbf{h}}^{(l)\star}
                   C_{\mathbf{0}}^{(j')} C_{\mathbf{g'}}^{(j')\star}
                   D_{\mathbf{0}}^{(l')\star} D_{\mathbf{h'}}^{(l')}
 \nonumber \\ & \times &
    e^{i(\gamma^{(l)}-\gamma^{(l')})t} \frac{1}{N_\mathbf{R}} \sum_\mathbf{R} e^{i(\mathbf{q}-\mathbf{q'})\cdot\mathbf{R}}
 \nonumber \\ & \times &
   \frac{1}{N_\mathbf{u}} \sum_\mathbf{u} \frac{S_\mathbf{u}(\mathbf{q},\mathbf{q'},E)}{q^2 q'^2} e^{i(\mathbf{q}-\mathbf{q'})\cdot\mathbf{u}}
\end{eqnarray}
where
\begin{eqnarray}
  \mathbf{q}  & = & \mathbf{k}_\mathrm{out} + \gamma^{(l)}\mathbf{\hat{n}}_\mathrm{out} - \mathbf{k}_\mathrm{in} - \gamma^{(j)}\mathbf{\hat{n}}_\mathrm{in} + \mathbf{h} - \mathbf{g} \\
  \mathbf{q'} & = & \mathbf{k}_\mathrm{out} + \gamma^{(l')}\mathbf{\hat{n}}_\mathrm{out} - \mathbf{k}_\mathrm{in} - \gamma^{(j')}\mathbf{\hat{n}}_\mathrm{in} + \mathbf{h'} - \mathbf{g'}
\end{eqnarray}
The Bloch coefficients $C_\mathbf{g}^{(j)}, D_\mathbf{h}^{(l)}$ for incoming and outgoing beam, respectively, are indexed by beams $\mathbf{g},\mathbf{h}$ and Bloch wave indices $j,l$. The elongations of the wave vectors perpendicular to the surface are denoted $\gamma^{(j)},\gamma^{(l)}$ for incoming and outgoing Bloch waves, respectively. The $t$ is the thickness of the crystal, $N_\mathbf{R}$ and $N_\mathbf{u}$ are number of unit cells and basis size. The mixed dynamical form-factor (MDFF) is denoted $S_\mathbf{u}(\mathbf{q},\mathbf{q'},E)$, where $\mathbf{q},\mathbf{q'}$ and $E$ are the momentum transfer vectors and energy loss, respectively. For more details about the theory of Bloch waves  we refer the reader to original literature \cite{rossouw,kohl,saldin} or a more recent literature\cite{prbtheory,bwconv} using the same notation as here.

A BW for incoming and outgoing beam can be expressed as
\begin{eqnarray}
  \psi_\mathrm{in}(\mathbf{r})  & = & \sum_{j\mathbf{g}} C_\mathbf{0}^{(j)\star} C_\mathbf{g}^{(j)} e^{i(\mathbf{k}_\mathrm{in} + \gamma^{(j)}\mathbf{\hat{n}}_\mathrm{in}+\mathbf{g})\cdot\mathbf{r}} \\
  \psi_\mathrm{out}(\mathbf{r}) & = & \sum_{l\mathbf{h}} D_\mathbf{0}^{(l)\star} D_\mathbf{h}^{(l)} e^{i(\mathbf{k}_\mathrm{out} + \gamma^{(l)}\mathbf{\hat{n}}_\mathrm{out}+\mathbf{h})\cdot\mathbf{r}} e^{i\gamma^{(l)}t}
\end{eqnarray}

The incoming wave can be formally written as
\begin{equation}
  \psi_\mathrm{in}(\mathbf{r}) = e^{i\mathbf{k}_\mathrm{in}\cdot\mathbf{r}} \sum_{g_x,g_y} F_{g_x,g_y}^{(z)} e^{i(g_x x + g_y y)}
\end{equation}
where
\begin{equation}
  F_{g_x,g_y}^{(z)} = \sum_{jg_z} C_\mathbf{0}^{(j)\star} C_\mathbf{g}^{(j)} e^{i\gamma^{(j)}z} e^{i g_z z}
\end{equation}
assuming that $\mathbf{\hat{n}}_\mathrm{in}=(0,0,1)$. Note that the $F_{g_x,g_y}^{(z)}$ is directly accessible from multislice calculation propagating the incoming beam, where $\{g_x,g_y\}$ correspond to a grid in the Fourier space.

We can accumulate the 2-dimensional arrays $F_{g_x,g_y}^{(z)}$ as a function of $z$, forming thus a 3-dimensional array with two dimensions in $g_x,g_y$ and third dimension in $z$. A Fourier transform with respect to the $z$-coordinate will provide a 3-dimensional array $F_\mathbf{\tilde{g}}$ using which we can draw the following parallels with Bloch waves theory
\begin{eqnarray}
  \psi_\mathrm{in}(\mathbf{r}) & = & e^{i\mathbf{k}_\mathrm{in}\cdot\mathbf{r}} \sum_{\mathbf{\tilde{g}}} F_\mathbf{\tilde{g}} e^{i\mathbf{\tilde{g}}\cdot\mathbf{r}} \\
  \mathbf{\tilde{g}} & \leftrightarrow & \mathbf{g} + \gamma^{(js)} \mathbf{\hat{n}}_\text{in} \label{eq:gmult}
\end{eqnarray}
finally allowing to write the following expression for DDSCS
\begin{eqnarray} \label{eq:dscsmix2}
  \frac{\partial^2 \sigma}{\partial\Omega \partial E} & = & 
    \sum_{\substack{\mathbf{hh'}ll'\\\mathbf{\tilde{g}\tilde{g}'}}} F_\mathbf{\tilde{g}}
                   D_{\mathbf{0}}^{(l)} D_{\mathbf{h}}^{(l)\star}
                   F_\mathbf{\tilde{g}'}^\star
                   D_{\mathbf{0}}^{(l')\star} D_{\mathbf{h'}}^{(l')}
 \nonumber \\ & \times &
    e^{i(\gamma^{(l)}-\gamma^{(l')})t} \frac{1}{N_\mathbf{R}} \sum_\mathbf{R} e^{i(\mathbf{q}-\mathbf{q'})\cdot\mathbf{R}}
 \nonumber \\ & \times &
   \frac{1}{N_\mathbf{u}} \sum_\mathbf{u} \frac{S_\mathbf{u}(\mathbf{q},\mathbf{q'},E)}{q^2 q'^2} e^{i(\mathbf{q}-\mathbf{q'})\cdot\mathbf{u}}
\end{eqnarray}
where
\begin{eqnarray}
  \mathbf{q}  & = & \mathbf{k}_\mathrm{out} + \gamma^{(l)}\mathbf{\hat{n}}_\mathrm{out} - \mathbf{k}_\mathrm{in} - \mathbf{h} - \mathbf{\tilde{g}} \\
  \mathbf{q'} & = & \mathbf{k}_\mathrm{out} + \gamma^{(l')}\mathbf{\hat{n}}_\mathrm{out} - \mathbf{k}_\mathrm{in} - \mathbf{h'} - \mathbf{\tilde{g}'}
\end{eqnarray}

\begin{figure}[thb]
  \includegraphics[width=7cm]{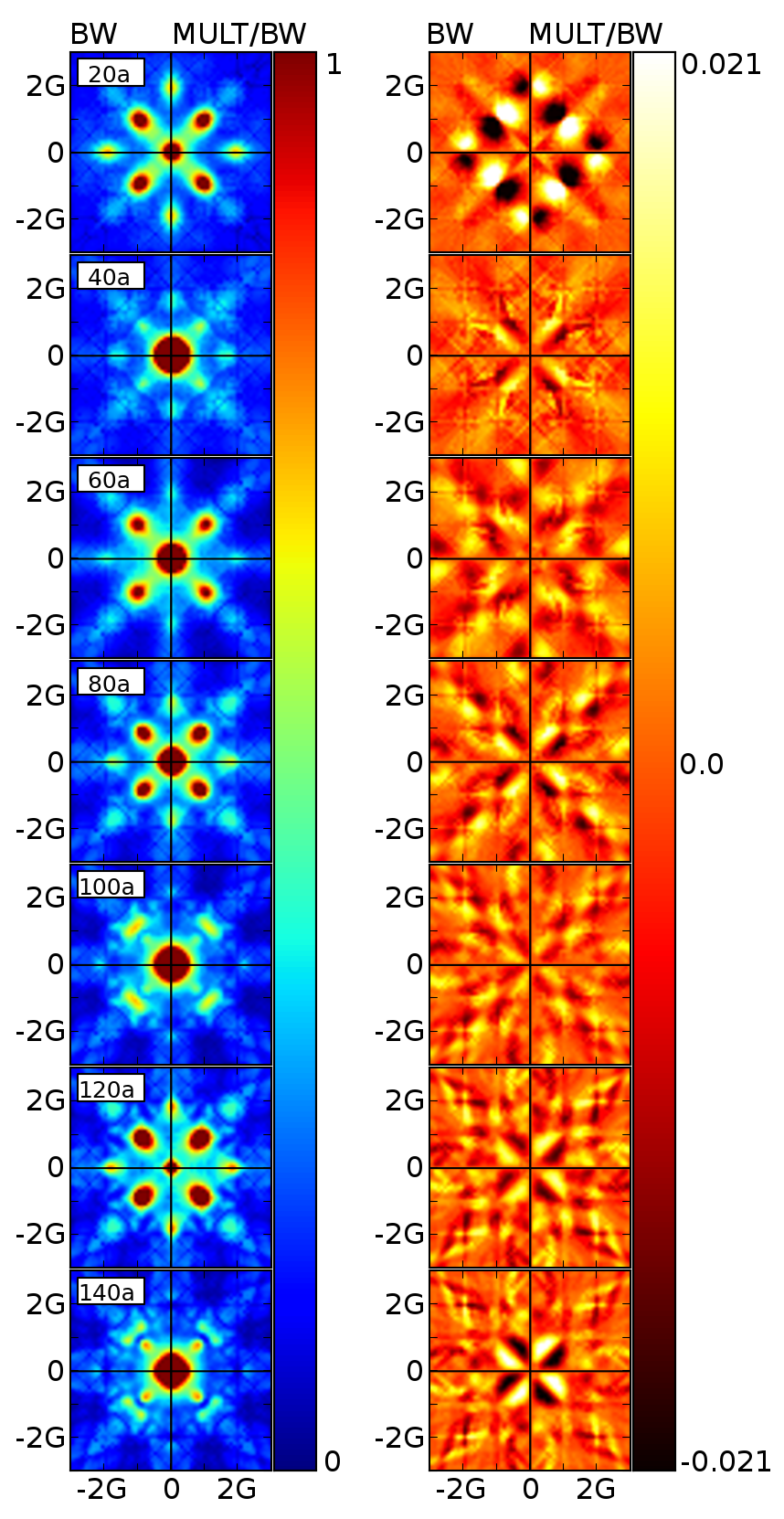}
  \caption{Calculation of inelastic scattering diffraction patterns (left) and distribution of the magnetic signal (right) for the $L_3$ edge of iron oriented along $(001)$ zone axis. Incoming beam is a plane wave at 200~keV, crystal thickness varies from $20a$ up to $140a$, where $a=2.87$~\AA{}. A Bloch waves (BW) calculation\cite{bwconv} is compared to a combined multislice/Bloch waves (MULT/BW) approach described in this manuscript.}
  \label{fig:planewave}
\end{figure}

Equation~(\ref{eq:gmult}) deserves a few remarks. The grid of $\tilde{g}_z$ depends on the slice thickness and on total thickness of the simulated column. The number of slices per unit cell limits the HOLZ contributions included in the calculation (maximum $\tilde{g}_z = 2\pi/\Delta z$). In our calculations, we used 42 slices per lattice parameter, but for generating the $F_\mathbf{\tilde{g}}$ we only used 6 slices per unit cell in the $z$-direction, thus maximal $\tilde{g}_z$ is $\frac{12\pi}{a}$. For the opposite limit, only wavelengths that are shorter than the column length can be recovered (minimum $\tilde{g}_z = 2\pi/t$). In our calculations we went up to approximately 40nm, but the fineness of the grid depends on the chosen thickness.

Having this in mind, the Bloch wave-vector elongations $\gamma^{(j)}$ are included via the $\tilde{g}_z$ values on the grid, as specified in the above-mentioned Eq.~(\ref{eq:gmult}). As a consequence, for a sufficiently large thickness this approach does not include any approximations and can fully recover the accuracy provided by Bloch waves method. This has been tested for a plane wave illumination and somewhat surprisingly, already for very thin specimen, both approaches provided very similar results, see Fig.~\ref{fig:planewave}.

Note further that vortex beam calculations simulated within a $20 \times 20$ supercell allow lateral components of the $\mathbf{\tilde{q}}$ to be fractions of the reciprocal lattice to the crystal unit cell. Summation over the lattice vectors $\mathbf{R}$ will however cancel out all terms for which $\mathbf{q}-\mathbf{q'}$ is not a reciprocal lattice vector of the unit cell. This is equivalent to the incoherent summation over the illumination angle, as long as the CBED disks do not overlap \cite{stembook,findlay}.

%Concerning the atomic-resolution scanning transmission electron microscopy (STEM) simulation we add a technical remark of importance when shifting the beam position. This is implemented in the multislice code via shifting the potential instead and keeping the beam in the middle of the simulation cell. That has to be taken into account when combining it with Bloch waves in Eq.~\ref{eq:dscsmix2} via the shift theorem of discrete Fourier transform, which introduces phase factors for the Fourier components of the incoming waveunction.

The formulation in Eq.~(\ref{eq:dscsmix2}) also lends itself for a straightforward modification of the \textsc{mats} summation algorithm\cite{bwconv}, where the products $C_\mathbf{0}^{(j)\star}C_\mathbf{g}^{(j)}$ are replaced by $F_\mathbf{\tilde{g}}$, and $\mathbf{g}$ and $\gamma^{(j)}$ are extracted from $\mathbf{\tilde{g}}$ according to Eq.~(\ref{eq:gmult}). The first step of calculating an energy-filtered diffraction pattern is a multislice propagation of an electron beam wavefunction. This is followed by a post-processing stage, where we extract the largest $F_\mathbf{\tilde{g}}$ and their corresponding $\mathbf{\tilde{g}}$ vectors. The $\mathbf{\tilde{g}}$ vectors are subsequently mapped on a pair of $\mathbf{g}$ and $\gamma^{(j)}\mathbf{\hat{n}}_\text{in}$. A modified version of the \textsc{mats} code loads the $F_\mathbf{\tilde{g}}$ and corresponding $\mathbf{g}$ and $\gamma^{(j)}\mathbf{\hat{n}}_\text{in}$ as a complete characteristics of the incoming beam wavefunction, and the rest (outgoing beam and summation) proceeds without any changes with respect to the original \textsc{mats} algorithm, only requiring that $xy$ components of  $\mathbf{q}-\mathbf{q'}$ correspond to a reciprocal lattice vector.

We argue that using multislice method for the incoming beam and Bloch waves method for outgoing beam is for our purpose an optimal combination of approaches because: 1) the complexity of the incoming beam would in BW method require a large number of independent calculations for all directions of incoming wave-vectors---and that is elegantly solved by a single run of the multislice method, and 2) after an inelastic event the propagation of electrons in various directions out of the sample can be naturally projected on BW fields corresponding to different outgoing plane-wave directions.

In the further subsections we apply this method to calculate the dependence of scattering of EVBs on magnetic properties of sample as a function of their diameter, initial angular momentum, acceleration voltage and displacement from atomic column. Influence of the detector shape is discussed (circular vs annular), as well as the relation between the angular momentum and channeling of EVB and its sensitivity to EMCD as a function of sample thickness. In these calculations we have used about 600 Bloch waves for the description of the outgoing beam and a summation cut-off criterion\cite{bwconv} of $P_\text{min}=10^{-5}$. The diffraction patterns were evaluated on a grid of $101 \times 101$ pixels spanning a range from $-5G$ to $5G$ with a step of $0.1G$ in both $k_{f,x},k_{f,y}$ directions, where $\mathbf{G}=(100)$.

\subsection{Detector shape considerations\label{sec:detshape}}

\begin{figure}[thb]
  \includegraphics[width=8.6cm]{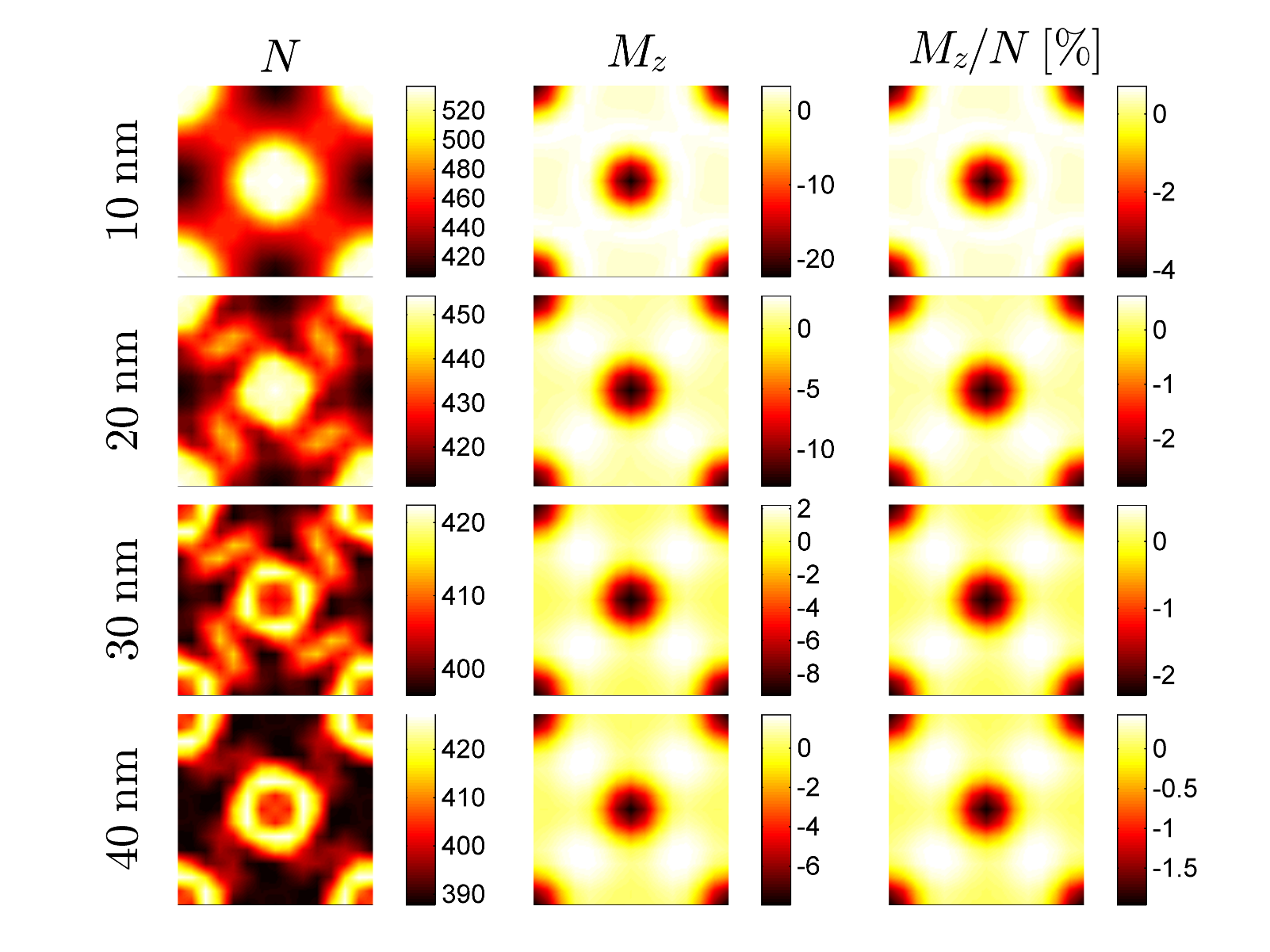}
  \caption{BF energy-filtered image of the non-magnetic signal (left column) and absolute magnetic signal (middle column) shown in arbitrary units, but on a common scale. Resulting relative magnetic signal (right column) is in percent. Calculations were performed with a circular aperture of radius $1.8\qmax$. Each map covers a single unit cell.}
  \label{fig:mapcirc15}
\end{figure}

\begin{figure}[thb]
  \includegraphics[width=8.6cm]{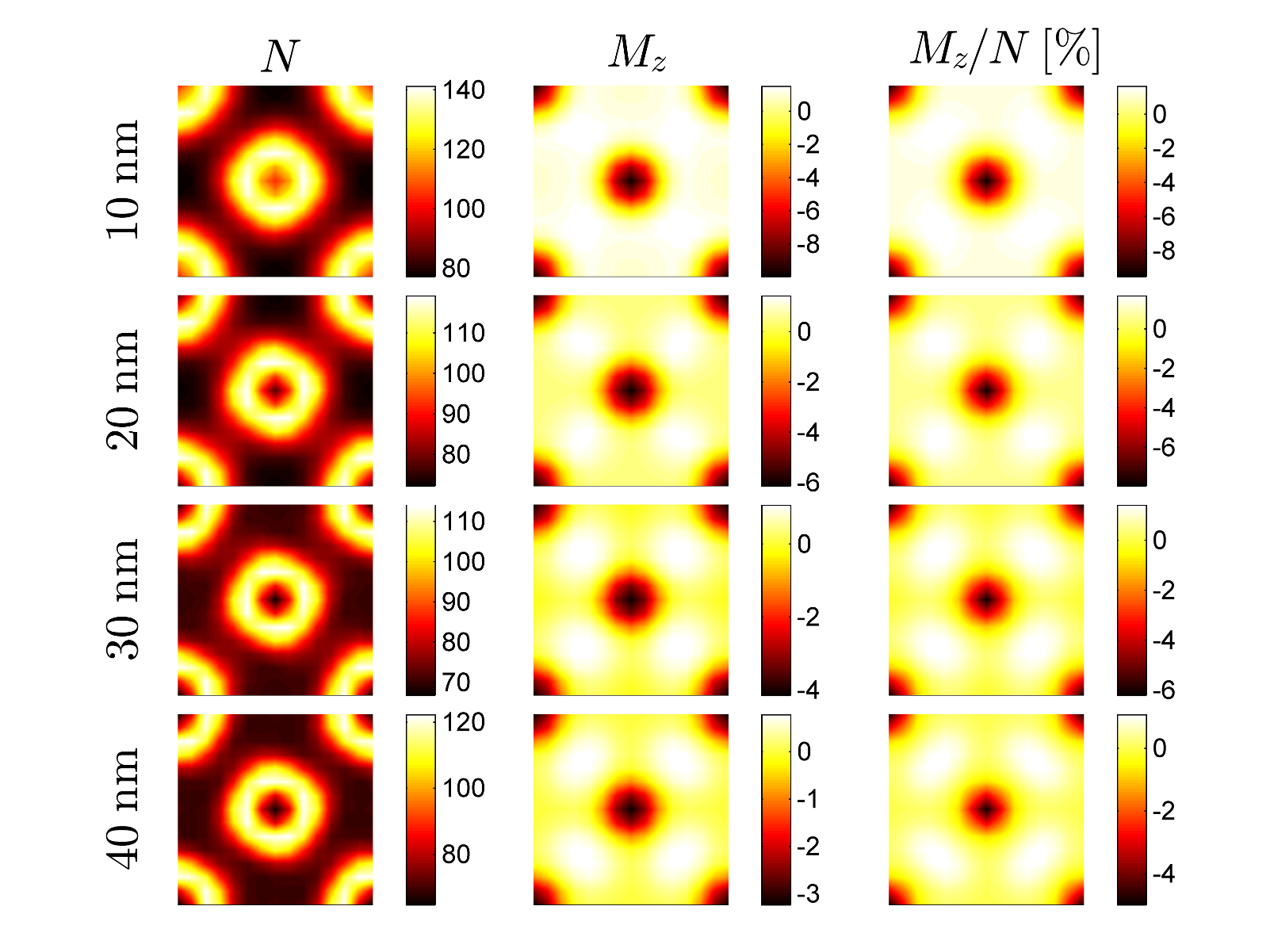}
  \caption{ADF energy-filtered image (see Fig.~\ref{fig:mapcirc15} for details) calculated using inner and outer detector radii of $1.2\qmax$ and $1.8\qmax$, respectively.}
  \label{fig:mapann1510}
\end{figure}
  
It was already mentioned that the potential advantage of EVBs in measurement of EMCD originates in the possibility to acquire data at the transmitted beam, which means much stronger signal compared to measurements in between Bragg spots, which is the case of intrinsic EMCD\cite{nature}. Nevertheless, it is not a priori obvious, what is the optimal shape of the detector for measurement of EMCD even if the beam is an EVB. In the case of strong vortex EMCD, the magnetic signal still varies in sign in the diffraction plane, typically forming concentric features. For EVBs with $\lz=\pm\hbar$ typically we can identify two regions: a narrow approximately circular area in the middle of diffraction plane with a magnetic signal having the same sign as the angular momentum, and a much broader annular region with an opposite sign of EMCD. Thus it may be of advantage not to measure the central spot in the diffraction plane, but rather to collect data over an annular area. These two options of signal collection, the circular and annular aperture centered on the transmitted beam, are realized in TEM by a bright field (BF) and annular dark field (ADF) detector, respectively. Corresponding two modes of operation in the STEM are BF and ADF imaging, respectively.

This is illustrated for $\qmax=0.5$~a.u.$^{-1}$, $V=200$~keV and $\lz=1\hbar$ in Figs.~\ref{fig:mapcirc15} and \ref{fig:mapann1510}. These figures show atomic-resolution energy-filtered images\cite{prange} of the non-magnetic signal ($N$) and absolute ($M_z$) and relative ($M_z/N$) magnitudes of the magnetic signal, respectively. Fig.~\ref{fig:mapcirc15} shows BF image for a detector aperture with diameter $1.8\qmax$ and Fig.~\ref{fig:mapann1510} shows ADF image for an aperture with inner and outer diameters of $1.2\qmax$ and $1.8\qmax$, respectively.

The aperture for the BF image, Fig.~\ref{fig:mapcirc15}, was chosen to contain the whole central region with the positive EMCD signal and a large part of the negative annular region with negative EMCD signal. Obviously, the magnetic contribution from the annular region dominates the magnetic signal at atomic columns. The nonmagnetic contribution shows non-trivial contrast variations in the image as a function of thickness. This originates from the inelastic scattering to the smallest angles. By choosing an annular detector aperture, Fig.~\ref{fig:mapann1510}, the nonmagnetic signal variations are substantially suppressed and we observe only donut-shaped features around the atomic columns. Such STEM image is obviously simpler to interpret in terms of positions of atoms. The maps of the magnetic signal show well localized peaks at atomic columns in both cases. For a small circular aperture encircling only the positive EMCD signal (not shown) the interpretation of the atomic resolution STEM image is complicated due to strong dynamical diffraction effects. The EMCD signal fraction has a complicated distribution as well and its relative strength does not go beyond 1\% and absolute strength is almost 20 times lower than for aperture shapes in Figs.~\ref{fig:mapcirc15} and \ref{fig:mapann1510}, thus a measurement should be focused on the large annular region of negative magnetic signal.

%Comparing the numerical values in these figures, we point out two observations regarding the strength of the magnetic signal. 1) In terms of absolute magnitude, the circular aperture should offer the strongest signal. In such case, it is of advantage to use a collection angle larger than the convergence angle. We have not concluded an optimal upper limit, but the magnetic signal becomes very weak above $1.8\qmax$, which is almost $5.0G$. 2) In terms of relative magnitude, large circular apertures are not optimal. The reason is the bright central spot, which contributes with a lot of intensity of nonmagnetic signal, though with little of EMCD, which is moreover of opposite sign compared to larger scattering angles. Thus a higher fractional EMCD signal can be obtained for an annular aperture with a larger outer radius and an inner radius, which at minimum blocks the area with magnetic signal of opposite sign. Optimizing the relative signal leads to unreasonably thin apertures, which would not be practical because of the very low signal that they would acquire. However, it is seen that a magnetic signal of relative strength of about 10\% can be acquired without substantially sacrificing the signal to noise ratio.

When processing data from the large parameter space survey, in most of the cases, where a sizable absolute EMCD signal could be detected, an optimal aperture seems to be an annular aperture with outer diameter of about $5G$ with $\mathbf{G}=(100)$ and inner diameter of about $0.9G$. Because of the dependence on the reciprocal lattice vectors, the collection angles will depend on the acceleration voltage. Otherwise, the dependence on other parameters ($\qmax$, $\lz$, sample thickness and beam displacement) appears to be rather weak. Note that $5G$ is the maximum radius fitting the range of the calculated diffraction patterns. Thus it is possible that a further increase of the total EMCD signal is achievable with larger apertures. However, based on the decay of the magnetic signal with larger scattering angles we do not expect any strong enhancement.

\subsection{Dependence of vortex-EMCD on beam diameter\label{sec:beamdiam}}

\begin{figure}[t!]
  \includegraphics[width=8.6cm]{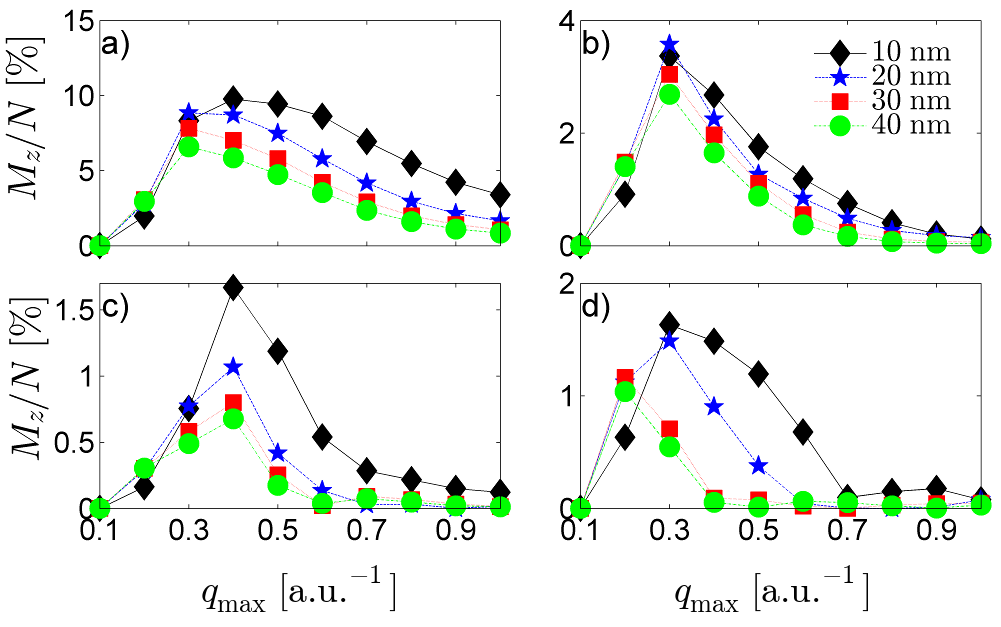}
  \includegraphics[width=8.6cm]{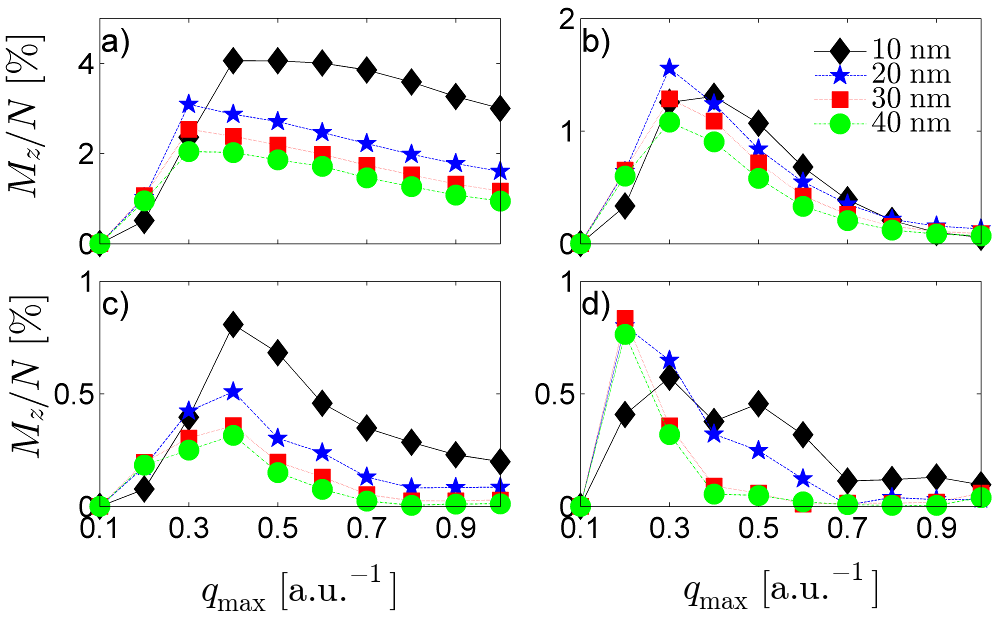}
  \includegraphics[width=8.6cm]{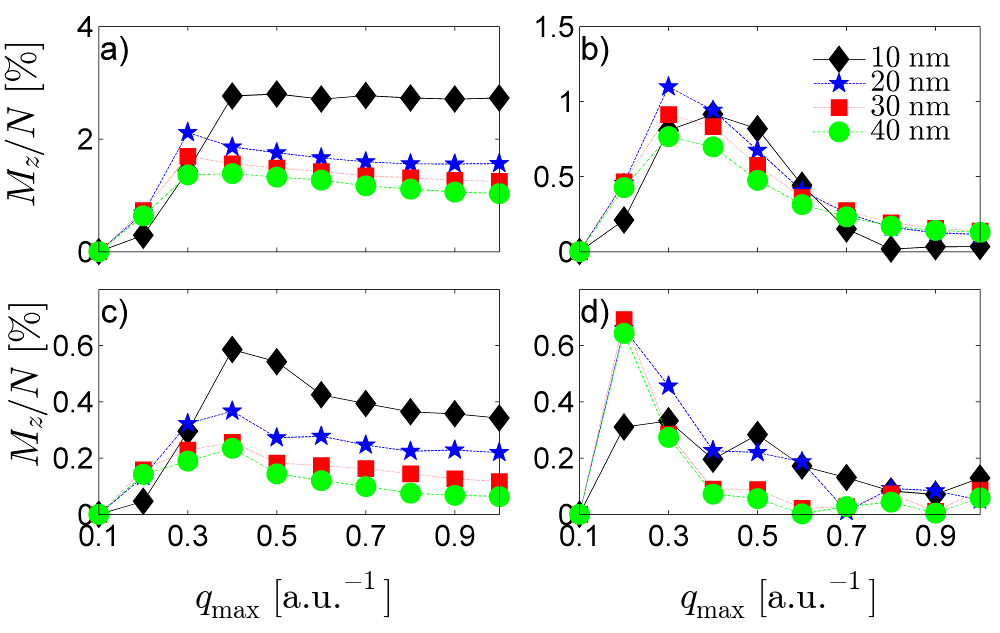}
  \caption{Relative strength of the EMCD signal as a function of beam diameter parametrized via $\qmax$. Acceleration voltage is 200~keV and $\lz=1\hbar$. The shift from the atomic column is a) 0~\AA, b) 0.41~\AA, c) 0.82~\AA, and d) 1.23~\AA, respectively. Shape of the aperture was (top) annular with diameters $0.6$~a.u.$^{-1}$ and $0.9$~a.u.$^{-1}$, (middle) circular with diameter $0.9$~a.u.$^{-1}$ and (bottom) circular with diameter $0.6$~a.u.$^{-1}$.}
  \label{fig:qmaxdep}
\end{figure}

%It was already demonstrated in subsection~\ref{sec:emcdonset} that at 200~keV the radius of the CBED disk must be larger than $\qmax \approx 0.13$~a.u.$^{-1}$ in order to observe vortex induced EMCD signal. 
Here we try to answer the question, what is the optimal probe size in terms of the relative strength of the EMCD signal. We have tested this for three types of detector apertures: two circular aperture diameters of $0.6$~a.u.$^{-1}=1.2\qmax$ and $0.9$~a.u.$^{-1}=1.8\qmax$, and an annular aperture that is defined as an area between these two radii, see Fig.~\ref{fig:qmaxdep}. The figure explores the relative strength of EMCD also as a function of sample thickness and displacement of the beam from the atomic column.

As a function of thickness, the relative strength of EMCD typically decreases. Exceptions are observed for the smaller circular aperture and occassionally for low $\qmax$ values. Similarly, displacing the beam from the atomic column mostly suppresses the EMCD signal. The signal is invariably strongest, when the beam passes directly through an atomic column. Even as small displacement as 0.41~\AA{} causes a drop of the signal strength at least by factor of two. For more details see the next subsection.

As a function of beam diameter, the EMCD signal fraction first increases with $\qmax$, then typically reaches a maximum between $\qmax=0.3$~a.u.$^{-1}$ and $0.5$~a.u.$^{-1}$. In most cases, the maximum is reached at $\qmax=0.4$~a.u.$^{-1}$. At settings used here ($\Vacc=200$~keV, $\lz=1\hbar$) that translates to an EVB with FWHM of $3.1$~$\text{a.u.}=1.6$~\AA. After reaching the maximum the relative strength of EMCD typically decays. I.e., much narrower EVBs do not necessarily help to improve detection of EMCD. A curious exception is the case of smaller circular aperture evaluated for sample thickness of 10~nm. In this case the relative signal reaches a plateau at $\qmax=0.4$~a.u.$^{-1}$ and stays approximately constant up to the largest $\qmax$ value considered in this study. The absolute EMCD strength is however decreasing.

\subsection{Systematic survey of the parameter space\label{sec:surveyres}}

\begin{figure}[htb]
  \includegraphics[width=8.6cm]{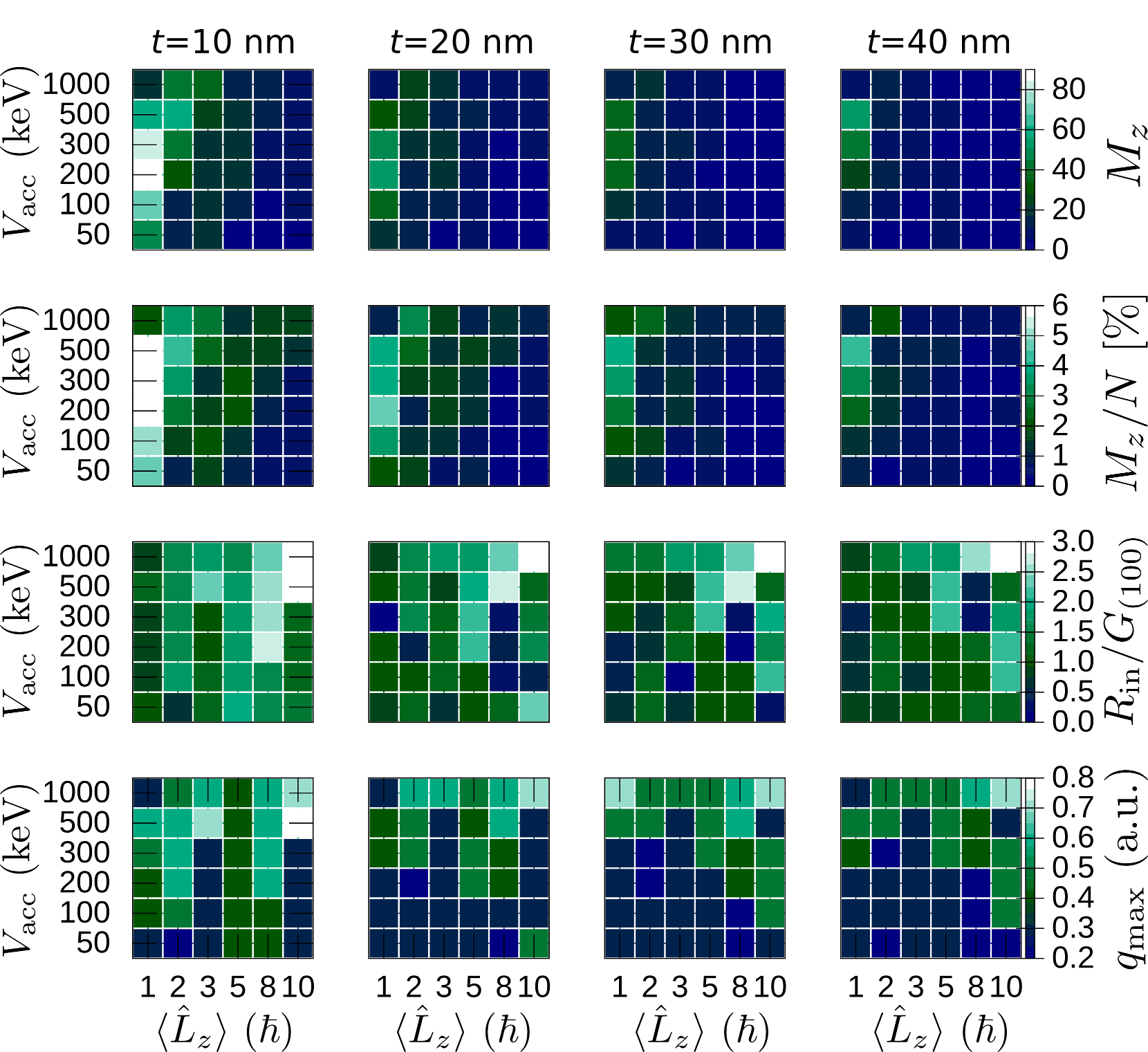}
  \caption{Optimized measurement conditions (beam diameter $\qmax$ and inner detector aperture $R_\text{in}$) and corresponding absolute $M_z$ and relative $M_z/N$ strength of the vortex-induced EMCD signal as a function of thickness $t$, acceleration voltage $\Vacc$ and OAM $\lz$.}
  \label{fig:survey}
\end{figure}

From the large parameter survey we have extracted the optimal $\qmax$ as a function of $\Vacc$ and $\lz$ and the results are summarized in the Fig.~\ref{fig:survey}. The image summarizes a rather large amount of data, so it deserves a detailed commentary. First of all, the data shown are for an EVB passing directly through an atomic column. For each combination of parameters an energy-filtered diffraction pattern has been calculated. The non-magnetic and magnetic contributions to the diffraction pattern were stored separately. For every such magnetic component of diffraction pattern we have calculated an optimized shape of the detector aperture by freely varying the inner and outer radii $R_\text{in}$ and $R_\text{out}$, respectively. Maximum value of $5G_{(100)}$ was allowed for these radii based on the range of calculated diffraction patterns. The criterion of optimization was to obtain a maximum absolute value of integrated magnetic signal. The outer radius of the detector aperture invariably reached the value of $R_\text{out}=5G$, which indicates that a larger outer aperture might still enhance the magnetic signal further. In the next step of processing the simulation data, we have chosen an optimal beam radius---this was parametrized by $\qmax$, see Table~\ref{tab:params}. The same criterion of optimization was chosen---an optimization of the absolute strength of magnetic signal. After this second step of optimization, the results were summarized and plotted in Fig.~\ref{fig:survey} as a function of thickness $t$, acceleration voltage $\Vacc$ and OAM $\lz$.

%The top row shows the absolute strength of the magnetic signal denoted $M_z$.
A number of findings can be concluded from this figure. The primary result is the absolute optimum of the whole survey for bcc iron oriented along $(001)$ zone axis: $\qmax=0.4$~a.u.$^{-1}$, $\Vacc=200$~keV, $\lz=1\hbar$, $t=10$~nm, $R_\text{out}=5G_{(100)}$ and $R_\text{in}=0.8G_{(100)}$. At these conditions we predict the strongest absolute magnetic signal, i.e., the highest count rate in an experiment. Considering values\cite{opmaps} for a spin moment of iron $m_s=2.3\mu_B$ and number of holes in the $3d$ shell $N_h=3.7$, the relative strength of the magnetic signal at the iron $L_3$ edge obtained as a difference of the $L_3$ edge signals measured with EVBs with OAM $+\hbar$ and $-\hbar$ divided by their average is about 7\%. We note that it is possible to obtain significantly higher relative magnetic signal fractions $M_z/N$, but at the cost of reduced electron counts. Since in the atomic-resolution STEM measurements of electron energy-loss spectra (EELS) have to cope with low electron count rates, optimization of the absolute electron count rates appears to be a more relevant criterion than a relative signal strength. Most suitable would be to optimize signal to noise ratio (SNR), but that depends also on the background signal under the core-level spectra, which we can't reliably estimate. If background signal dominates the intensity, then the core-level spectrum is only a small modification of the total electron count at given energy loss. In such case, the absolute strength of magnetic signal is practically proportional to the SNR, because the noise is mostly determined by slowly varying background.

It has been suggested that EVBs with a large starting angular momentum could be more efficient in detecting EMCD\cite{mcmorran,saitoh2}. Our simulations demonstrate that for lower or medium acceleration voltages it is most efficient to measure with EVB with OAM $\pm\hbar$. Only at $\Vacc=500$~keV or more, occassionally we can obtain stronger magnetic signal with $\lz=\pm2\hbar$. This advantage however reduces and eventually disappears with increasing thickness $t$. At $\Vacc=1000$~keV the strength of magnetic signal obtainable with $\lz=3\hbar$ is very close to the maximum value for $\lz=2\hbar$, which suggests that at even larger acceleration voltages it might be advantageous to work with EVBs with $\lz=3\hbar$ or even more. Though for practical purposes, in the range of the most common acceleration voltages it appears that a choice of $\lz=\pm\hbar$ is optimal.

The optimum at the lowest sample thickness, as mentioned above, is obtained for $\Vacc=200$~keV. However, for larger thicknesses, the optimal acceleration voltage is expectedly moving to higher $\Vacc$. At $t=30$~nm an optimal acceleration voltage is $\Vacc=300$~keV and at $t=40$~nm it moves up to $\Vacc=500$~keV. Even at these conditions, the absolute magnetic signal strength is quite considerable---being about 60\% of the overal optimum. Thus if a characterization of thicker samples is desired, simulations suggest to use higher $\Vacc$ and, as can be inspected from the bottom row of Fig.~\ref{fig:survey}, a beam with a smaller diameter (that is larger $\qmax$).

The evolution of the optimum $\qmax$ and $R_\text{in}$ is less systematic as a function of the other parameters. This is probably a consequence of complicated dynamical diffraction effects involved in the process of inelastic scattering of EVBs. Generally, for the cases where a reasonably strong magnetic signal can be observed, the $\qmax$ typically stays within a range of $0.2$--$0.6$~a.u.$^{-1}$ and the $R_\text{in}$ is in the range from $0.5G_{(100)}$ to $2.0G_{(100)}$.

\subsection{Displacement of the EVB from atomic column}

In the previous subsection we have considered an EVB passing directly through a column of atoms. In our survey we have also considered a range of shifts of the EVB core from the atomic column. The smallest considered shift $\datt=0.14$~\AA{} causes a little of qualitative change. The obtainable strength of the magnetic signal is reduced mostly by 5--10\% and occassionally up to 20\%. The optimum conditions remain unchanged, only the strength of the signal is reduced by 9.5\%. Generally, we conclude that misplacing the beam witin $\pm 0.14$~\AA{} from the atomic column causes only a minor weakening of magnetic signal.

Displacing the beam further to $\datt=0.34$~\AA{} still shows the same qualitative trends, however with significantly reduced strength of the magnetic signal. In most cases the reduction falls into an interval 20--45\%. %At the same time we observe that that usefullness of EVBs with larger OAM is slowly suppressed. 
Yet, the magnetic signal keeps the same sign. Thus a measurement of EMCD with EVBs within in a radius of $\pm 0.35$~\AA{} from the atomic column should still provide a signal with a sizable magnitude.

Situation qualitatively changes at a shift of $\datt=0.68$~\AA{}. The magnetic signal drops to values that reach not more than 12\% of the optimum. The reduction of the magnetic signal seems to be less strong for beams with larger OAM. As a consequence, at this displacement, EVBs with $\lz=2\hbar$ or even $3\hbar$ can provide strongest magnetic signal, especially at medium or lower voltages and higher sample thicknesses. But it should be stressed that here we talk about generally very weak magnetic signal, which especially in terms of relative magnitude barely reaches 1\% of the non-magnetic signal. In this context, our simulations suggest in order to measure an appreciable magnetic signal, positioning of the EVB core needs to be within a distance well below $0.68$~\AA{}.

Finally, we have performed calculations for an EVB placed into the center of the edge of unit cell, i.e., displaced by $a/2$ from all nearest atomic columns ($a=2.87$~\AA{} is the unit cell parameter of bcc iron). The magnetic signal detectable here is somewhat stronger, its magnitude reaches up to 30\% of the optimum, but its sign is opposite to the sign of magnetic signal nearby atomic columns. However, a relative strength of the magnetic signal is not larger, it is only about 2--3\%. %As a result, in an experiment it will may be difficult to resolve such small signal component.

\begin{figure}[thb]
  \includegraphics[width=8.6cm]{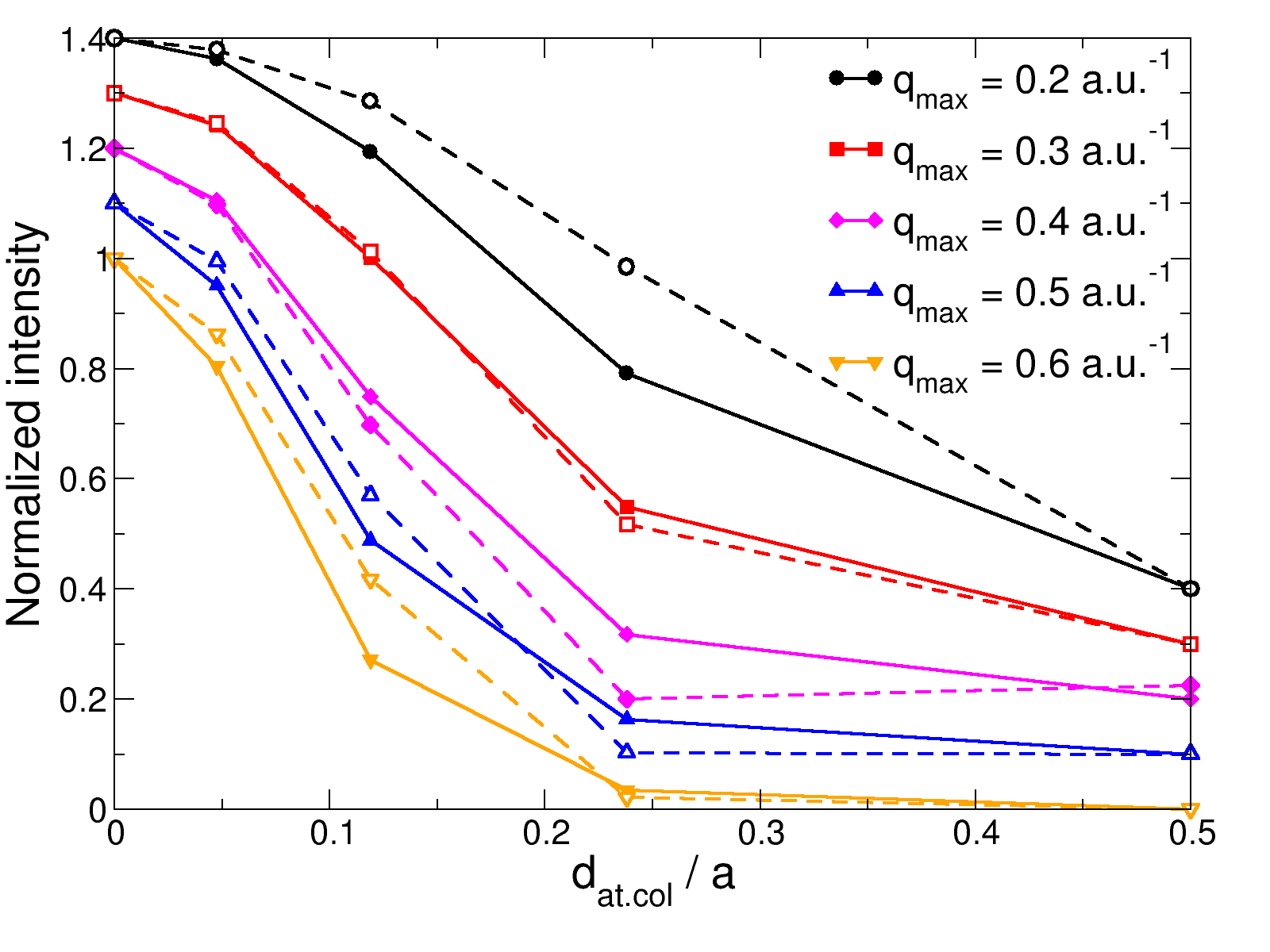}
  \caption{Intensity of the nonmagnetic signal for $\lz=0$ (full lines) and magnetic signal for $\lz=-\hbar$ (dashed lines) as a function of distance from atomic column $\datt$. An annular detector as in Fig.~\ref{fig:mapann1510} was used. The intensity was normalized to the range from 0 to 1 for all dependences and a vertical offset was used for clarity.}
  \label{fig:fwhmnhmz}
\end{figure}

We conclude this subsection with a curious observation. In the optimum conditions, that is $\Vacc=200$~keV, $\lz=1\hbar$ and $\qmax=0.4$~a.u.$^{-1}$ the magnetic signal drops to its half at a distance slightly beyond $0.34$~\AA{}. That means that the magnetic signal is very localized close to the atomic column, giving it an effective FWHM of approximately $0.7$~\AA{}. What makes this observation rather unexpected is that the beam itself has a FWHM of $1.6$~\AA{}, i.e., substantially larger. Even more surprising it becomes, when we realize that an ordinary convergent beam with zero OAM produced at the same conditions has a FWHM of $0.71$~\AA{}. In other words, it appears that a magnetic signal measured by an EVB with $\lz=1\hbar$ provides very similar spatial resolution as a beam with zero OAM in detection of the non-magnetic signal. The same observation can be made from the energy-filtered high-resolution maps calculated for $\qmax=0.5$~a.u.$^{-1}$ published in Ref.~\cite{vortexelnes}. Thus it is appealing to hypothesize that it could be a systematic phenomenon. That would mean that an EVB with $\lz=1\hbar$ offers the same spatial resolution in magnetic signal as is achievable by an ordinary beam ($\lz=0$) prepared in the same conditions for a usual STEM-EELS. We illustrate this observation more explicitly in Fig.~\ref{fig:fwhmnhmz}, where an intensity profile of a non-magnetic signal for beam with zero OAM is compared to an intensity profile of magnetic signal measured with EVB having $\lz=-1\hbar$. To put them on the same intensity scale, the profiles were rescaled to the range from 0 to 1. As one can see, for all considered $\qmax$ values the FWHM of the magnetic signal profile is close to the FWHM of nonmagnetic signal, despite that the EVB is more than twice as wide as the ordinary beam, Fig.~\ref{fig:fwhmxqmax}. At present we can't offer any qualitative explanation of this finding and postpone this question for a further investigation.

\subsection{Relation between vortex EMCD and angular momentum of the beam and channeling\label{sec:chan}}

\begin{figure}[thb]
  \includegraphics[width=8.6cm]{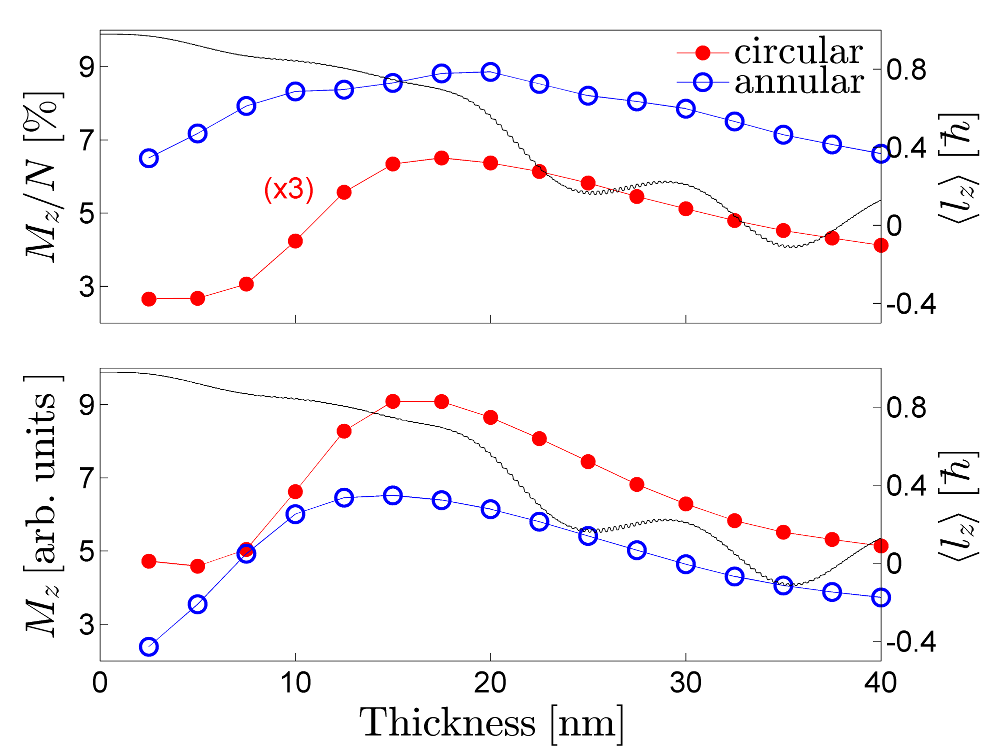}
  \caption{The top panel in this figure displays the thickness dependence of the relative signal, $M_z/N$, obtained with circular and annular detectors, as in Figs.~\ref{fig:mapcirc15} and \ref{fig:mapann1510}. The magnitude of the signal from the circular detector is multiplied by three for convenience reasons. The lower panel is showing the magnitude of the magnetic signal in $z$-direction. Both panels also have a reference curve of the thickness dependence of the angular momentum associated with the right axis.}
  \label{fig:lzvsemcd}
\end{figure}

The exchange of angular momentum between the sample and EVB has been studied in Ref.~\cite{vortcryst} and recently in Ref.~\cite{vortexelnes}. It is tempting to suggest that there should be a link between the ability of the beam to transfer the angular momentum to the sample and its sensitivity to the magnetic properties of the sample. Though one can also argue against by noting that the exchange of the angular momentum happens also in the elastic regime, where sample acts as an infinite reservoir and beam changes its angular momentum without noticeable energy loss. Moreover, as was already shown in Ref.~\cite{vortexelnes}, even a wide EVB exchanges the angular momentum with the sample, despite that its sensitivity to the magnetic properties of sample is negligible. Here we probe explicitly this question by performing a simulation of thickness dependence of the angular momentum of the EVB and a thickness dependence of the (relative and absolute) strength of the detected magnetic signal using both circular and annular apertures. The results are summarized in Fig.~\ref{fig:lzvsemcd}. As can be seen, there is no visible correlation between the two properties, other than a more-or-less decreasing trend with oscillations at thicknesses above 20nm. That reiterates the need for an explicit calculation of the inelastic scattering effects, when attempting to quantify the sensitivity of EVB to materials magnetic properties.

\begin{figure}[t!]
  \includegraphics[width=8.4cm]{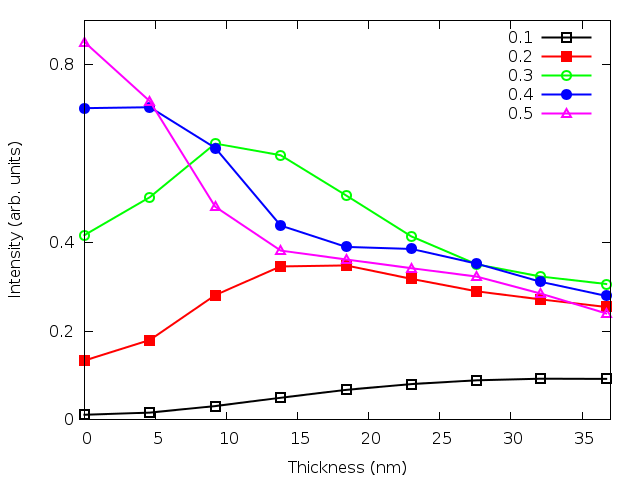} 
  \includegraphics[width=8.4cm]{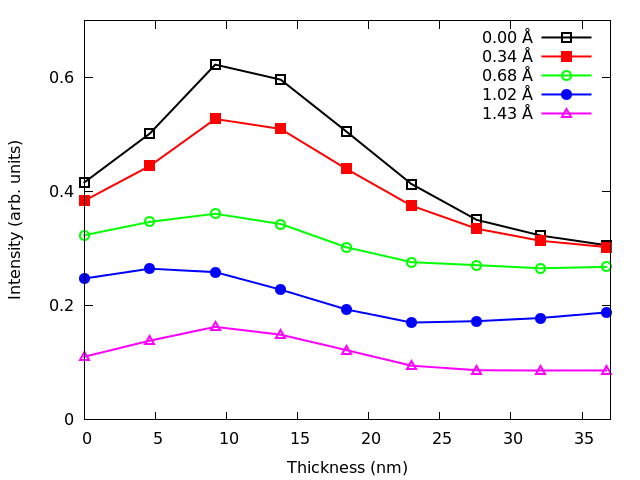} 
  \includegraphics[width=8.4cm]{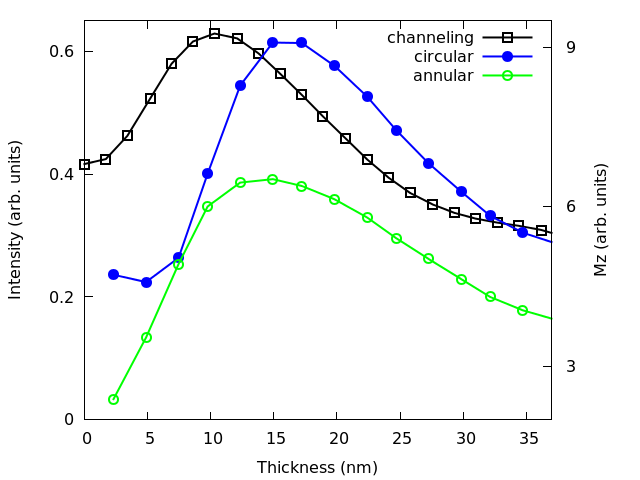} 
  \caption{Channeling strength per atom as a function of thickness: a) for beams with different diameter (parameterized by $\qmax$) centered on an atomic column and (b) for $\qmax{}=0.3$~a.u.$^{-1}$, varying the distance of beam center from an atomic column; (c) a comparison between the EMCD strength and channeling strength as a function of thickness for a beam of $\qmax{}=0.3$~a.u.$^{-1}$ and centered on an atomic column. The beam energy is taken to be 200~keV in all the figures.}
  \label{fig:channeling}
\end{figure}

Channeling is another parameter, that can be estimated directly from elastic scattering calculations and it would be interesting to investigate whether it is correlated to the EMCD strength. As an electon beam enters the crystal, it is attracted towards the
positively charged atomic columns. As a result, the beam tend pass through the crystal along channels formed around the atomic columns. We define channeling strength per atomic column as the integrated intensity of electron beam within 1~\AA{} diameter around an atomic column (distance of closest atomic columns in bcc iron is $\frac{a}{2}\sqrt{2}\approx 2.03$~\AA{}). We illustrate channeling as a function of beam diameter and position of beam center in Fig.~\ref{fig:channeling}a and \ref{fig:channeling}b, respectively. Clearly, as the beam diameter is reduced, channeling per atomic column increases, because wider beams tend to channel through a larger number of atomic columns than a narrow beam. However, beyond $\qmax{}=0.3$~a.u.$^{-1}$ channeling decays rapidly as the beam passes through the crystal, most likely due to fast spreading of very narrow beams. Placing the beam center away from the atomic column also reduces the chanelling strength (see Fig.~\ref{fig:channeling}b. Qualitatively, these two observations are in accord with calculated EMCD strength dependences. Encouraged by this, we directly compared the chanelling strength (for a 200 keV beam, with $\qmax{}=0.3$~a.u.$^{-1}$, centered on an atomic column) to an EMCD strength in Fig.~\ref{fig:channeling}c. Although there is some qualitative similarity (e.g., chanelling strength peaking around 10--12~nm thickness and then decaying, similar to the EMCD strength), however, it is not an one to one correlation and an explicit calculation of EMCD effect appears to be the only safe option for estimating its strength dependence.

\section{Conclusions\label{sec:concl}}

We have systematically explored the elastic and inelastic scattering of the EVBs on a bcc iron crystal as a traditional benchmark system for studies of EMCD.
%Our results suggest that the EMCD caused by EVBs is an interference effect and, as such, it onsets when the CBED disks start overlapping in the diffraction plane. At an acceleration of 200~keV that happens for an EVB with $\lz=1\hbar$ at relatively large FWHM of 5.0~\AA. Yet, the onset of vortex EMCD is weak and optimal detection is achieved for substantially narrower beams.

Optimization of the detector aperture shape suggests to use an annular aperture with an outer radius of $5G$ or more [$\mathbf{G}=(100)$] and an inner radius of approximately $G$, depending on the angular momentum of the beam and, to some extent, also on acceleration voltage. The strongest absolute EMCD signal for a 10~nm layer of bcc iron was observed for EVB passing directly through a column of atoms and having the following parameters: $\lz=1\hbar$, acceleration voltage 200~keV, and $\qmax=0.4$~a.u.$^{-1}$, which corresponds to a FWHM of 1.6~\AA. Under these conditions, the EMCD signal constitutes about 7\% of the white-line intensity at the iron $L_3$ edge. In terms of relative EMCD strength it is possible to obtain higher signal fractions (over 10\%), but at the cost of overal lower signal strength, thus sacrificing the measured electron count rate.

EMCD signal appears to be well localized nearby the atomic columns and quickly decreases with increasing distance of the vortex core from the atomic column. In the optimal measurement conditions stated above, the EMCD strength drops to about half of its maximum value at a distance of approximately $0.35$~\AA{} leading to an atomic resolution map of the EMCD signal, which has a FWHM of about $0.7$~\AA{} around atomic columns. It is worth noting that this is much less than the FWHM of the vortex beam itself, while it is surprisingly close to the FWHM of a beam with $\lz=0$ obtained in the same conditions. Reasons for this unexpected yet systematic observation are not understood at this point.

Increase of the beam angular momentum in majority of cases does not lead to an increase of the EMCD strength. On the other hand, the beams with larger angular momentum appear to be somewhat less sensitive to a slight displacement of the beam from the atomic column. Still, in terms of the maximum strength of EMCD signal, optimal value remains $\lz=1\hbar$.

For an EVB with $\lz=1\hbar$ the acceleration voltage has, similarly as the beam diameter, a non-monotonous influence on the EMCD strenth. As stated above, an optimum is reached around 200~keV. For EVBs with higher starting $\lz$ the strength of EMCD slowly increases with acceleration voltage, mostly monotonously. Yet, even at 1000~keV, the absolute strength of EMCD does not overcome the optimal value for $\lz=1\hbar$ at 200~keV. Thus, increasing of the acceleration voltage is not likely to bring a major advantage in measuring EMCD with vortex beams, except for thicker samples: at 40~nm, the optimum settings for EVB with $\lz=1\hbar$ are acceleration voltage of 500~keV and $\qmax=0.5$~a.u.$^{-1}$, which is a vortex beam with FWHM of 1.3~\AA. Compared to the 10~nm thickness the loss of the optimal signal strength it about 35\%, i.e., vortex beams may offer a rather efficient route also for thicker samples.

Finally, we have found no clear correlations between the $z$-dependent angular momentum of the probe and the sensitivity to the magnetic properties. On the other hand, we found some similarity between the electron channeling and strength of EMCD, though it is not a clear correspondence. Hence, to obtain an estimate of the EMCD strength, one has to perform an inelastic electron scattering calculation.

In this work we have not considered an influence of aberrations on the sensitivity of EVBs to magnetic properties. Without doubts it is an important aspect for the feasibility of such measurements. On the other hand, the optimal settings require relatively narrow probe sizes, which are mostly achieved in aberration corrected microscopes, where the defocus and spherical aberration can be efficiently suppressed. Source size broadening was also not considered here. In terms of high-resolution energy-filtered STEM images this will cause a smearing of the contrast and, because the intensity of the magnetic signal rapidly drops with displacement of the EVB from atomic column, the optimal absolute and relative magnitudes of the EMCD strength will be correspondingly reduced.

In summary, detection of EMCD with EVBs should be feasible with instruments available today\cite{krivanek}, assuming that spectra with sufficient signal to noise ratio can be obtained.

\section{Acknowledgements}

Inspiring discussions with Shunsuke Muto, Kazuyoshi Tatsumi, Koh Saitoh and Nobuo Tanaka are gratefully acknowledged. J.R. acknowledges Swedish Research Council, G\"{o}ran Gustafsson's Foundation, Swedish National Infrastructure for Computing (NSC center) and computer cluster \textsc{Dorje} at Czech Academy of Sciences.

\end{document}